\documentclass[a4paper,11pt]{article}

\usepackage{amsmath,amssymb}
\usepackage{bm}
\usepackage{upgreek}
\usepackage{url}
\usepackage{adjustbox}
\usepackage{setspace}
\usepackage{textcomp}
\usepackage{rotating}
\usepackage{gensymb}
\usepackage[numbers]{natbib}
\usepackage[english]{babel}
\usepackage[T1]{fontenc}
\usepackage{float}
\usepackage[top=1in, bottom=1.25in, left=1.25in, right=1.25in]{geometry}
\usepackage{url}

\makeatletter
\let\@fnsymbol\@arabic
\makeatother

\title{Emulation of CPU-demanding reactive transport models: comparison of Gaussian processes, polynomial chaos expansion and deep neural networks}

\author{Eric Laloy\thanks{Belgian Nuclear Research Centre. Email: {\tt elaloy@sckcen.be}} and Diederik Jacques$^1$}
\begin{document}

\maketitle

\begin{abstract}
	
This paper presents a detailed comparison between 3 methods for emulating CPU-intensive reactive transport models (RTMs): Gaussian processes (GPs), polynomial chaos expansion (PCE) and deep neural networks (DNNs). State-of-the-art open-source libraries are used for each emulation method while the CPU-time incurred by one forward run of the considered RTMs varies from 1h to between 1h30 and 5 days. Besides direct emulation of the simulated uranium concentration time series, replacing the original RTM by its emulator is also investigated for global sensitivity analysis (GSA), uncertainty propagation and probabilistic calibration using Markov chain Monte Carlo (MCMC) sampling. The selected DNN is found to be superior to both GPs and PCE in reproducing the input - output behavior of the considered 8-dimensional and 13-dimensional CPU-intensive RTMs. This even though the used training sets are small: from 75 to 500 samples. Furthermore, the two used PCE variants: standard PCE and sparse PCE (sPCE) appear to always provide the least accuracy while not differing much in performance. As a consequence of its better emulation capabilities, the DNN method outperforms the two other methods for uncertainty propagation. Regarding GSA, the DNN and GP methods offer equally good approximations to the true first-order and total-order Sobol' sensitivity indices while PCE does less well. Most surprisingly, despite its superior emulation skills the DNN approach leads to the worst solution of the considered synthetic inverse problem which involves 1224 measurement data with low noise. This apparently contradicting behavior of the used DNN is at least partially due to the small but complicated deterministic noise that affects the DNN-based predictions. Indeed, this complex error structure can drive the emulated solutions far away from the true posterior distribution in case of high quality measurement data. Among the considered 3 methods only the GP method allows for retrieving emulated posterior solutions that jointly (1) fit the high-quality measurement data to the appropriate noise level (log-likelihood value) and (2) most closely fit the true model parameter values. Overall, our findings indicate that when the available training set is relatively small (75 - to 500 input - output examples) and fixed beforehand, PCE is not the best choice for emulating CPU-intensive RTMs. Instead GPs or DNNs should be preferred. However, a DNN can deliver overly biased model calibration results. In contrast, the GP method performs fairly well across all considered tasks: direct emulation, global sensitivity analysis, uncertainty propagation, and calibration.

\end{abstract}

\section{Introduction}
\label{intro}
Simulating the fate and transport behavior of radionuclides and other reactive solutes in the vadose zone and aquifers requires reactive transport models \citep[RTMs, e.g.,][]{Steefel2005,Steefel2015}. These RTMs can be quite CPU-demanding, especially when simulating interactions between multiple species along with dynamic sorption and decay processes based on thermodynamic and kinetic laws \citep[][]{Jacques2008,Liu2008, Yin2011,Greskowiak2010,Greskowiak2015, Jacques2018}. This type of RTM is often termed ``multicomponent'' RTM. Given that the computational cost incurred by a single RTM run can vary from a few minutes for simple RTMs to days for multicomponent RTMs, any task that necessitates many RTM runs may benefit from the construction of an emulator or ``surrogate'' model. The latter is basically a mathematical function that aims to reproduce the input - output behavior of the physically-based model, and evaluating this function for a given input set has negligible or very small computational requirements. When devising the emulator does not require any modification to the original physically-based model, the emulator is said to be ``non-intrusive''. Hence, non-intrusive emulation considers the original model as a black-box. Also, emulators can rely on simplified physics or be purely statistical. This work is rooted within the most convenient framework of non-intrusive statistical emulation.

As for any other environmental model, simulation-based decision making with a RTM typically entails uncertainty propagation (UP). The UP process consists of randomly sampling the model input parameters (i.e., the model parameters sensu stricto, possibly including the initial and/or boundary conditions as well) from their joint probability density function (pdf) and running the model for each draw such that statistics of the resulting model response can be computed. Prior to UP, the modeler may want to perform a sensitivity analysis (SA) to establish to what extent each input parameter influences one or more model outputs that are deemed important for the task at hand. For comprehensive sensitivity information, global sensitivity analysis \citep[GSA, e.g.,][]{Iooss-Lemaitre2015} that screens the whole parameter space and accounts for parameter interactions is to be preferred over a so-called ``local'' SA which provides sensitivity indices that depend on the selected starting point and ignore parameter interactions. In addition, when measurement data are available the joint pdf of the considered model input parameters can be estimated within a Bayesian framework by probabilistic inverse modeling (calibration). This requires using sampling methods such as, e.g., Markov chain Monte Carlo \citep[MCMC, e.g.,][]{Robert-Casella2004} sampling, while the inferred parameter distribution is called the posterior parameter pdf.

The 3 important tasks of UP, GSA and inversion all involve repeated model runs. Depending on the considered task and model complexity, from say 1000 to 1,000,000 (or more) runs may be needed \citep[e.g.,][]{Laloy-Vrugt2012, Laloy2015, Marrel2015,Roy2018}. This can make GSA, UP and, especially, inversion intractable for RTMs. In this work, we investigate the use of 3 methods to emulate RTMs, thereby facilitating GSA, UP and inversion. Each of these tasks has its own requirements in term of accuracy of the emulator. For GSA one does not necessarily need a large accuracy but only that the sensitivity of the emulator outputs to the input parameters is similar to that of the original model's outputs. A more accurate emulator is needed for UP in the sense that the relevant statistics of the emulator's output must be close to those of the original model's output. Lastly, emulation-based inverse modeling requires by far the largest accuracy of the emulator. Indeed, very slight differences in response between the emulator and the original model may cause proportionally large deviations between the emulated and true posterior parameter distributions.

For non-intrusive emulation of RTMs, we consider herein the following 3 state-of-the-art nonlinear regression methods. The most recently developed method is a deep neural network \citep[DNN, e.g.,][]{Goodfellow2016} of the classical feed-forward fully connected (FC). To our knowledge this is the first application of deep learning (DL) to the emulation of RTMs. Our DNNs are constructed using the open-source Keras \citep{Keras2015} Python package which is built on top of the open-source Tensorflow DL library \citep{tensorflow2016}. The two other methods are perhaps the most popular emulation techniques in engineering and the geosciences: Gaussian processes \citep[GPs, ][]{Rasmussen-Williams2006} and polynomial chaos expansion \citep[PCE, see, e.g.,][just to list a few]{Ghanem-Spanos1991,Sudret2008,Blatman-Sudret2010,Lemaitre-Knio2010,Shao2017}. In both cases we use state-of-the-art open source implementations: the GPflow Python package \citep{Matthews2017} for GPs, which is built on top of Tensorflow, and the Chaospy Python package \citep{Feinberg-Langtangen2015} for PCE. Importantly, we do not consider adaptive or ``goal-oriented'' emulator construction in this study. Instead, we focus on the rather common situation where the training set is fixed and built beforehand.

Our two considered modeling case studies involve simulation of the 8- and 13-dimensional multi-rate surface complexation and 1D transport of Uranium VI (U(VI)) in a soil column over time \citep[][]{Greskowiak2015}. These two problems are established benchmarks to test the accuracy of reactive transport codes \citep[][]{Steefel2015,Greskowiak2015}. While the first problem considers a single domain with constant hydrochemical composition of the inflow water, the second problem includes dual domain mass transfer and alternating hydrochemical composition at the column inlet. The associated CPU-times for a single RTM run are on the order of 1h and 1h30 to 5 days, respectively. In addition, this second problem is assumed to cover all important processes that are deemed to control uranium transport under real field conditions \citep[][]{Greskowiak2015}.

In summary, this study evaluates three techniques (DNNs, GPs, and PCE) in their ability to emulate two related RTM problems and to facilitate three different tasks (UP, GSA, inversion). The remainder of this paper is organized as follows. Section \ref{related_work} skims through the related work before section \ref{methods} describes the proposed DNN and used GP and PCE emulation methods. Section \ref{prob_description} then presents the considered 2 modeling case studies, and section \ref{bench} lays out the main principles behind GSA, UP and inversion. This is followed by section \ref{results} which details for the two case studies the performance of DNNs, GPs and PCE for direct emulation, GSA, UP and inversion (first modeling case study only). In section \ref{discussion}, we recap our main results, provide recommendations, and outline possible future DNN developments. Finally, section \ref{conclusion} concludes with a summary of our most important findings.

\section{Related work}
\label{related_work}
Although the resulting benefits are potentially huge in terms of computational savings, emulation of RTMs has been scarcely addressed so far. \citet{Jatnieks2016} present a short comparison of a range of nonlinear regression methods for the emulation of the geochemical component of a relatively simple RTM. Some important differences with our work are as follows. Firstly, Jatnieks and coworkers are concerned with replacing the geochemical component of their RTM by an emulator trained beforehand while assuming the geochemical model parameters to be fixed. Since geochemical calculations commonly take most of the running time of an RTM, this can be very useful when the goal is to perform one or more otherwise CPU-intensive simulation(s) with fixed model parameters. Nevertheless, this strategy is not suited to tasks involving varying the model parameter values such as GSA, UP and inversion. As stated earlier, we emulate the full RTM response to (some) model parameter and boundary condition changes. This classical emulation strategy permits emulator-based GSA, UP and inversion and has the advantage of being non-intrusive: no modifications to the RTM need to be made. Note that in principle, one could plug an emulator of the geochemical solver within an RTM and try to build a new emulator on the fly for each evaluated model parameter set within a GSA, UP or inversion procedure. The technical feasibility and tractability of such approach remain to be investigated though. Our RTM problems are also more complex than that considered by \citet{Jatnieks2016}, with equilibrium speciation and kinetics, more than 10 aqueous components, 100 (problem 1) to 200 (problem 2) kinetic reactions and mixing of different boundary solutions \citep[problem 2, see][for details]{Greskowiak2015}. The work by \citet{Sun2012} reports on the calibration of 4 first-order reaction rates using a rather simple RTM that does not account for thermodynamic equilibrium, inter-species interactions and sorption processes. In addition, Sun and coworkers do not mention which emulation method they use exactly. \citet{Keating2016} resort to multivariate adaptive regression splines to emulate a model of the multi-phase transport of CO$_2$ with non-reactive transport of associated solutes. \citet{Keating2016} use their emulator for an UP application that consists of predicting the ensemble behavior of some given model outputs across the considered parameter space.

\section{Emulation Methods}
\label{methods}
Here we provide a brief description of the considered 3 nonlinear regression methods. The following terminology is employed
	\begin{equation}
	\textbf{y} = M\left(\textbf{x}\right),
	\label{original_model}
	\end{equation}
and
	\begin{equation}
	\textbf{y}_s = S\left(\textbf{x}\right),
	\label{surrogate_model}
	\end{equation}
where $M\left(\cdot\right)$ is the original RTM, $S\left(\cdot\right)$ denotes its emulator or surrogate model, $\textbf{x}$ signifies the vector of model input parameters, and $\textbf{y}$ and $\textbf{y}_s$ represent the vector of original and emulated outputs, respectively.

\subsection{Gaussian Processes}
\label{methods_gp}

Gaussian processes (GPs) can be thought as generalization of a Gaussian probability distribution to functions. The core idea is that the original model or function to be emulated, $M\left(\textbf{x}\right)$, resembles a Gaussian stochastic process $G\left(\cdot,\cdot\right)$, which is completely determined by its mean function, $m\left(\textbf{x}\right)$, and covariance function, $k\left(\textbf{x},\textbf{x}'\right)$, \citep[see the reference book by][for details]{Rasmussen-Williams2006}
	\begin{equation}
	M\left(\textbf{x}\right) \approx S_{\rm GP \it}\left(\textbf{x}\right) = G\left(m\left(\textbf{x}\right),k\left(\textbf{x},\textbf{x}'\right)\right).
	\label{gp1}
	\end{equation}

Let us represent the ensembles of training and test parameter points by the $n \times d_x$ $\textbf{X}$ and $n^* \times d_x$ $\textbf{X}^*$ arrays, respectively, with $n$ the number of training instances, $n^*$ the number of test instances to be predicted and $d_x$ the dimensionality of $\textbf{x}$ (from 8 to 13 herein, see further). Similarly, the corresponding ensembles of training and yet to be predicted test outputs are denoted by the $n \times d_y$ $\textbf{Y}$ and $n^* \times d_y$ $\textbf{Y}^*_s$ arrays, respectively, with $d_y$ the dimensionality of $\textbf{y}$ (1224 herein, see further). It can be shown than the predictive distribution of a given column, $j$ (i.e., output at a given time step herein), of $\textbf{Y}^*_s$, (where for clarity, the ``s'' underscore has been dropped)
\begin{equation}
\textbf{Y}^*_{\cdot,j} | \textbf{X},\textbf{X}^*,\textbf{Y}_{\cdot,j} \propto N\left(\mathbb{E}\left[\textbf{Y}^*_{\cdot,j} | \textbf{X},\textbf{X}^*,\textbf{Y}_{\cdot,j}\right],\rm cov \it \left(\textbf{Y}^*_{\cdot,j}\right)\right)
\label{gp2}
\end{equation}
is given by \citep[][]{Rasmussen-Williams2006}
	\begin{equation}
	\mathbb{E}\left[\textbf{Y}^*_{\cdot,j} | \textbf{X},\textbf{X}^*,\textbf{Y}_{\cdot,j}\right] = m_j\left(\textbf{X}^*\right) + k\left(\textbf{X}^*,\textbf{X}\right)\left[k\left(\textbf{X},\textbf{X}\right)+\sigma^2_y\textbf{I}\right]^{-1}\left(\textbf{Y}_{\cdot,j} -  m_j\left(\textbf{X}\right)\right)
	\label{gp3}
	\end{equation}
	\begin{equation}
	\rm cov \it \left(\textbf{Y}^*_{\cdot,j}\right) =  k\left(\textbf{X}^*,\textbf{X}^*\right)- k\left(\textbf{X}^*,\textbf{X}\right)\left[k\left(\textbf{X},\textbf{X}\right)+\sigma^2_y\textbf{I}\right]^{-1}k\left(\textbf{X},\textbf{X}^*\right)
	\label{gp4}
	\end{equation}
where $k\left(\textbf{X}^*,\textbf{X}\right)$ is the $n^* \times n$ matrix of covariances between the test and training input points, $k\left(\textbf{X},\textbf{X}\right)$ is the $n \times n$ matrix of covariances between the training input points, $m_j\left(\cdot\right)$ is the mean function for the $j^{\rm th}$ output and $\sigma^2_y$ is the variance of a zero-mean white Gaussian noise that is associated with the measurement of $\textbf{y}$. Since we deal with deterministic computer simulations, in our case $\sigma^2_y = 0$. For the prediction of a single scalar output, $y^*_{i,j}$, equations (\ref{gp3}-\ref{gp4}) then reduce to
	\begin{equation}
	\mathbb{E}\left[y^*_{i,j} | \textbf{X},\textbf{x}^*,\textbf{Y}_{\cdot,j}\right] = m_j\left(\textbf{x}^*\right) + k\left(\textbf{x}^*,\textbf{X}\right)k\left(\textbf{X},\textbf{X}\right)^{-1}\left(\textbf{Y}_{\cdot,j} -  m_j\left(\textbf{X}\right)\right)
	\label{gp5}
	\end{equation}
	\begin{equation}
	\mathbb{V}\left[y^*_{i,j} | \textbf{X},\textbf{x}^*,\textbf{Y}_{\cdot,j}\right] = k\left(\textbf{x}^*,\textbf{x}^*\right)-k\left(\textbf{x}^*,\textbf{X}\right)k\left(\textbf{X},\textbf{X}\right)^{-1}k\left(\textbf{x}^*,\textbf{X}\right)^{\rm T}
	\label{gp6}
	\end{equation}
where $k\left(\textbf{x}^*,\textbf{X}\right)$ signifies the $1 \times n$ vector of covariances between the single test input point, $\textbf{x}^*$, and the $n$ training input points in $\textbf{X}$.

There are many possibilities for the choice of the covariance kernel  $k\left(\cdot,\cdot\right)$. We tested with every kernel available in GPflow (see \url{http://gpflow.readthedocs.io/en/master/}) and retained the following two ones, that we used either separately or as an additive combination (see further): the very popular squared exponential or Gaussian kernel (SE) and the second-order arccosine kernel (ACOS2). The SE can be written as
	\begin{equation}
	k\left(\textbf{x},\textbf{x}'\right) = \sigma^2_k\rm exp \it \left[-\frac{1}{2}\sum_{i=1}^{d_x}\left(\frac{\left(x_{i}-x_{i}'\right)}{l_i}\right)^2\right]
	\label{gp7}
	\end{equation}
where $\sigma^2_k$ and $\textbf{l} = \left[l_1,\cdots,l_{d_x}\right]$ are fitting parameters. The ACOS2 has a more complex form and is described in \citet[][]{Cho-Saul2009}. This kernel has $d_x$ + 2 fitting parameters.

With respect to the mean function, $m\left(\textbf{x}\right)$, a linear function is used
	\begin{equation}
	m\left(\textbf{x}\right) = \textbf{A}^{\rm T}\textbf{x}+\textbf{b}
	\label{gp8}
	\end{equation}
where $\textbf{A}$ is $d_x \times d_y$, $\textbf{x}$ is $d_x \times 1$ and $\textbf{b}$ is $d_y \times 1$. Similarly as for the kernel parameters (as in e.g., equation \ref{gp7}), the coefficients in $\textbf{A}$ and $\textbf{b}$ need to be estimated from the data. The $m\left(\cdot\right)$ and $k\left(\cdot,\cdot\right)$ parameters can be jointly estimated by maximum likelihood estimation \citep[see][]{Rasmussen-Williams2006}. To do so, GPflow can be interfaced with the Scipy \citep[][]{Jones2001} optimization library (see \url{http://gpflow.readthedocs.io/en/master/}).

\subsection{Polynomial Chaos Expansion}
\label{methods_pce}

Polynomial chaos expansion (PCE) approximates a given function (CPU-demanding model herein) by a weighted sum of orthogonal polynomials. The parameters in these polynomials are transformations of the parameters of the original CPU-demanding model. The PCE has been firstly introduced by \citet{Ghanem-Spanos1991} and has nowadays gained widespread use in engineering and the geosciences where it is still actively researched.

In the following, we describe the PCE expansion of equation (\ref{original_model}) for $d_y = 1$. If $d_y > 1$, the approximation is applied separately to each component of $\textbf{y}$. The $Q$-degree PCE approximation of $M\left(\textbf{x} \right)$ can be expressed as
	\begin{equation}
	S_{\rm PCE \it}\left(\textbf{x}\right) = \sum_{j = 0}^{U-1} \alpha_{j} \bm{\Uppsi}_{j}\left( \textbf{x}\right),
	\label{pce1}
	\end{equation}
where the deterministic coefficients $\alpha_{0}, \ldots, a_{U-1}$ are unknown, and $U$ denotes the total number of $d_x$-dimensional orthogonal polynomials $\bm{\Uppsi}_{j}\left( \textbf{x}\right)$ of degree not exceeding $Q$ in a total-order expansion. The value of $U$ is simply computed as
	\begin{equation}
	U = \frac{\left(Q + d_x\right)!}{Q! d_x!}.
	\label{pce2}
	\end{equation}
The $\bm{\Uppsi}_{j}\left( \textbf{x}\right)$ are products of the monodimensional polynomials for expansion terms $j = 1, \ldots, U$: $\bm{\Uppsi}_{j}\left( \textbf{x}\right) = \psi_{j,1}\left(x_1 \right) \times \psi_{j,2}\left(x_2 \right), \ldots, \times \psi_{j,n}\left(x_{d_x} \right)$.

Various polynomial types can be used in equation (\ref{pce1}) depending on the available prior information about the variables in $\textbf{x}$. For instance, Jacobi polynomials are typically associated with a beta distribution, Hermite polynomials with a Gaussian distribution, Laguerre polynomials with a gamma distribution, and Legendre polynomials, that represent a special case of the Jacobi polynomials, with an uniform distribution \citep{Lemaitre-Knio2010}. Different polynomial bases can also be combined. In this study, Legendre polynomials are used since we assume uniform distributions for the components of $\textbf{x}$ (see further).

There are two main non-intrusive methodologies to estimate the  $\alpha_{0}, \ldots, a_{U-1}$ coefficients. The so-called ``pseudo-spectral'' approach and least square (LS) minimization. In short, the former resorts to an numerical integration scheme to compute the coefficients while the latter solves a linear regression system. The pseudo-spectral approach thus requires the training points to be carefully selected, e.g, from a quadrature rule or a sparse grid \citep[see, e.g.,][for details]{Lemaitre-Knio2010}. In contrast, the LS approach can be used with any set of training points. Both pseudo-spectral and LS estimation are available in Chaospy \citep[see][and \url{http://chaospy.readthedocs.io/en/master/tutorial.html}]{Feinberg-Langtangen2015}. In principle, the LS method necessitates an over-determined set of of linear equations ($n_x$ > $U$)
	\begin{equation}
	\underbrace{\left[ \begin{array}{ccc}
		\bm{\Uppsi}_{0}\left( \textbf{x}_1\right) & \cdots & \bm{\Uppsi}_{U-1}\left( \textbf{x}_1\right)\\
		\vdots & \ddots & \vdots \\
		\bm{\Uppsi}_{0}\left( \textbf{x}_{n_x}\right) & \cdots & \bm{\Uppsi}_{U-1}\left( \textbf{x}_{n_x}\right)
		\end{array} \right]}_{\textbf{P}}
	\underbrace{\left[ \begin{array}{c}
		\alpha_{0} \\
		\vdots \\
		\alpha_{U-1}
		\end{array} \right]}_{\bm{\upalpha}}
	= \underbrace{\left[ \begin{array}{c}
		M\left(\textbf{x}_1\right) \\
		\vdots \\
		M\left(\textbf{x}_{n_x}\right)
		\end{array} \right]}_{\textbf{z}},
	\label{pce3}
	\end{equation}
and
	\begin{equation}
	\bm{\upalpha}=\left(\textbf{P}^{\rm T}\textbf{P}\right)^{-1}\textbf{P}^{\rm T}\textbf{z},
	\label{pce4}
	\end{equation}
where $\bm{\upalpha}$ is now the vector of estimated coefficients. In case the system in equation (\ref{pce3}) is rank-deficient or ill-conditioned (e.g., when $n_x$ < $U$), it is however better to use the singular value decomposition (SVD) method \citep[see, e.g.,][]{Aster2012} than equation (\ref{pce4}) directly. The SVD method is the default in the Chaospy package. Chaospy also features Tikhonov regularization \citep[as well as many other regularization strategies, see][]{Feinberg-Langtangen2015}, which gives rise to the following loss function to be minimized
	\begin{equation}
	\mathcal{L}_{\rm PCE \it} = ||M\left(\textbf{x}\right) - S_{\rm PCE \it}\left(\textbf{x}\right)||^2 + \lambda||\bm{\upalpha}||^\tau,
	\label{pce5}
	\end{equation}
where the $\lambda$ value balances the effect of the regularizatinon term on $\mathcal{L}_{\rm PCE}$ and is jointly estimated with $\bm{\upalpha}$ by generalized cross validation \citep[see][]{Feinberg-Langtangen2015}, and $\tau$ is the regularization order that is set to 2 herein. When Tikhonov regularization is used, the resulting PCE is called a ``sparse'' PCE (sPCE). Note that sPCEs can also be obtained from other regularization techniques \citep[e.g.,][]{Blatman-Sudret2010,Shao2017}.

\subsection{Deep Neural Networks}
\label{methods_dnn}

Neural networks basically define the (possibly complex) relationships existing between input, $\textbf{x}$, and output, $\textbf{y}$, data vectors by using combinations of computational units that are called neurons. A neuron is an operator of the form:
	\begin{equation}
	h\left(\textbf{x}\right) =f\left(\textbf{w}^{\rm T}\textbf{x} + b \right),
	\label{dnn1}
	\end{equation}
where $h\left(\cdot \right)$ denotes the scalar output of the neuron, $f\left(\cdot \right)$ is a nonlinear function that is called the ``activation function", $\textbf{w} = \left[w_1, \cdots, w_{d_x}\right]$ is a set of weights and $b$ represents the bias associated with the neuron. For a given task, the values for $\textbf{w}$ and $b$ associated with each neuron must be optimized or ``learned" such that the resulting neural network performs as well as possible. When $f\left(\cdot \right)$ is differentiable, $\textbf{w}$ and $b$ can be learned by gradient descent. Common forms of $f\left(\cdot \right)$ include the rectified linear unit (ReLU), sigmoid function and hyperbolic tangent function.

When there is no directed loops or cycles across neurons or combinations thereof, the network is said to be feedforward (FFN). In the FFN architecture, the neurons are organized in layers. A standard layer is given by
	\begin{equation}
	\textbf{h}\left(\textbf{x}\right)=f\left(\textbf{W}\textbf{x} + \textbf{b} \right),
	\label{dnn2}
	\end{equation}
where $\textbf{W}$ and $\textbf{b}$ are now a matrix of weights and a vector of biases, respectively. The name multilayer perceptron (MLP) designates a FFN with more than one layer. A most typical network is the 2-layer MLP, which consists of two layers with the outputs of the first-layer neurons becoming inputs to the second-layer neurons
	\begin{equation}
	\textbf{y}_s=\textbf{g}\left[\textbf{h}\left(\textbf{x}\right)\right]  \equiv f_2\left[\textbf{W}_2\cdot f_1\left(\textbf{W}_1\textbf{x} + \textbf{b}_1 \right) + \textbf{b}_2 \right],
	\label{dnn3}
	\end{equation}
where $\textbf{g}\left(\cdot\right)$ and $\textbf{h}\left(\cdot\right)$ are referred to as output layer and hidden layer, respectively.

In theory, the two-layer MLP described in equation (\ref{dnn3}) is a universal approximator as it can approximate any underlying process between $\textbf{y}$ and $\textbf{x}$ \citep{Cybenko1989,Hornik1991}. However, this only works if the dimension of $\textbf{h}\left(\cdot\right)$ is (potentially many orders of magnitudes) larger than that of the input $\textbf{x}$, thereby making learning practically infeasible. Instead, researchers have found that it is much more efficient to use many hidden layers rather than increasing the size of a single hidden layer \citep[e.g.,][]{Goodfellow2016}. When a FFN/MLP has more than one hidden layer it is considered to be deep. Nevertheless, current deep networks are not necessarily purely FFN but may mix different aspects of FFN, such as convolutional neural networks (CNN) and recurrent neural networks \citep[RNN, see, e.g.,][]{Goodfellow2016}.

After limited trial-and-error using problem 1 only (see section \ref{res_pbm1}), a 4-layer DNN, $S_{\rm DNN \it}\left(\textbf{x}\right)$, was selected for both problems 1 and 2 (see section \ref{prob_description}) with the following numbers of neurons for the $\textbf{h}_1\left(\cdot\right),\cdots,\textbf{h}_4\left(\cdot\right)$ layers: 64, 124, 256 and 1224, respectively. The hidden layers , $\textbf{h}_1\left(\cdot\right),\cdots,\textbf{h}_3\left(\cdot\right)$, are equipped with ReLU activation functions ( $f\left(x \right) = \max\left(0,x\right)$) while the output layer, $\textbf{h}_4\left(\cdot\right)$, has a linear (i.,e., identity) activation. The resulting total number of network parameters (weights vector and bias associated with every neuron) is 356,488 for problem 1 (where $\textbf{x}$ is 8-dimensional) and 356,808 for problem 2 (where $\textbf{x}$ is 13-dimensional, see section \ref{prob_description}).

To learn the network parameters, the following loss function with (partial) l2-norm or second-order Tikhonov regularization of the network weights is minimized
	\begin{equation}
	\mathcal{L}_{\rm DNN \it} = \left\|M\left(\textbf{x}\right) - S_{\rm DNN \it}\left(\textbf{x}\right)\right\|^2 + \gamma_1\left\|\textbf{W}_{1,2}\right\|^2 + \gamma_2\left\|\textbf{W}_{3}\right\|^2
	\label{dnn4}
	\end{equation}
where $\textbf{W}_{1,2}$ contains the weights associated with hidden layers $\textbf{h}_1\left(\cdot\right)$ and $\textbf{h}_2\left(\cdot\right)$, $\textbf{W}_{3}$ encodes the weights that belong to $\textbf{h}_3\left(\cdot\right)$ and the $\gamma_1$ and $\gamma_2$ regularization parameters are set to $1 \times 10^{-5}$ and $1 \times 10^{-8}$, respectively. The formulation in equation (\ref{dnn4}) was also chosen after limited trial-and-error using problem 1 only (see section \ref{res_pbm1}).

Since every used activation function is differentiable, equation (\ref{dnn4}) can be minimized by stochastic gradient descent (that is, gradient descent using a series of mini-batches rather than all the data at once) together with back propagation. This means that the loss function derivative is propagated backwards throughout the network using the chain rule, in order to update the parameters. Various stochastic gradient descent algorithms are available. In this work, we used the adaptive moment estimation (ADAM) algorithm which has been proven efficient for different types of deep networks \citep{Kingma-Ba2015}.

\section{Reactive Transport Problems}
\label{prob_description}
The considered reactive transport modeling problems are the first and last problems elaborated by \citet[][]{Greskowiak2015}. We refer herein to these two problems as problem 1 (and RTM1) and problem 2 (and RTM2). Greskowiak and coworkers compare several codes in their ability to simulate multi-rate surface complexation and 1D dual-domain multi-component reactive transport of U(VI). Problems 1 and 2 were crafted on the basis of previous studies investigating the desorption of U(VI) from radionuclide-contaminated sediment from the Hanford 300A site, Washington, USA \citep[][]{Greskowiak2015}.

\subsection{Problem 1}
\label{prob_1}
Problem 1 consists of the multi-rate surface complexation and 1D transport of U(VI) in a single domain with constant hydrochemical composition at the soil column inlet. The geochemical model simulates kinetic sorption of uranyl as two surface complexes: >SOUO2OH and >SOUO$_2$HCO$_3$ with equilibrium constants $K1$ and $K2$, respectively (>SOH represents a surface site for uranyl adsorption). Due to chemical and/or physical heterogeneity, a lognormal distribution of rate constants is considered to describe the kinetic sorption at different adsorptions sites. The geochemical model also considers equilibrium aqueous complexation with interactions of U with OH-, CO$_3$$^{2-}$, and some other elements. Therefore, the fate of U will be influenced by the pore water composition, pH and the partial pressure of CO$_2$ in the water, $p$CO$_2$ $\rm \left[-\right]$. The inflow alternates between prescribed flow and no-flow conditions in order to separate the effect of kinetic desorption from advection dispersion and to reproduce the impact of highly transient groundwater flow \citep[see][for details]{Greskowiak2015}. The corresponding RTM1 was implemented within the HPx software \citep{Jacques2018}. The 8 (varying) model parameters are \citep[see][for details]{Greskowiak2015}: the mean and standard deviation of the lognormal distribution of sorption rates, $\mu_r \left[\rm log\left(h^{-1}\right)\right]$ and $\sigma_r \left[\rm log\left(h^{-1}\right)\right]$, respectively, the logarithm of equilibrium constant $K1$, $logK1 \rm \left[-\right]$, the logarithm of equilibrium constant $K2$, $logK2 \rm \left[-\right]$, the total bulk sorption site density, $S^{\rm bulk}_{\rm tot}$, $\rm \left[mol/L_{bulk}\right]$, the total bulk initial uranium concentration, $U^{\rm bulk}_{\rm tot} \rm \left[mol/L_{bulk}\right]$, $p$CO$_2$ $\rm \left[-\right]$, and $pH$ $\rm \left[-\right]$, the pH of the solution. Bounds of the uniform parameter distribution are listed in Table \ref{table1}.

The HPx code supports Open-MP parallelization and when parallelized over 4 cores, one forward run takes about 1h on the used workstation. Lastly, among all the processes that are simulated, the quantity of interest in this work is the times series of dissolved U(VI) concentration (mol/l) in the outflow node of the mobile domain. This times series contains 1224 time steps. 

\subsection{Problem 2}
\label{prob_2}
Problem 2 adds complexity to problem 1. First, it further includes dual domain mass transfer in order to account for the field relevant effects of physical heterogeneity, (i.e., large centimeter to decimeter scale variations in permeability). Second, it also includes alternating hydrochemical composition at the column inlet. This result into a transient redistribution of uranium between the adsorbed and aqueous phase depending on aqueous major ion chemistry \citep[][]{Greskowiak2015}. As stated earlier, this problem encompasses the most relevant hydro(geo)logical and hydrochemical processes that influence uranium transport under real field conditions \citep[][]{Greskowiak2015}. The resulting RTM2 was also built within HPx. There are now 13 uniformly distributed free model parameters, of which lower and upper bounds can be found in Table \ref{table1}. Four parameters are similar as for problem 1: $\mu_r \left[\rm log\left(h^{-1}\right)\right]$, $\sigma_r \left[\rm log\left(h^{-1}\right)\right]$, $logK1 \rm \left[-\right]$, and $logK2 \rm \left[-\right]$. The 7 other free parameters are \citep[see][for details]{Greskowiak2015}: the total bulk sorption site density for the mobile domain $S^{\rm bulk}_{\rm tot,mob}$ $\rm \left[mol/L_{bulk}\right]$, the ratio of total bulk sorption site density for the immobile domain to $S^{\rm bulk}_{\rm tot,mob}$, $ratioS$ $\rm \left[-\right]$, the total bulk initial uranium concentration in the mobile domain, $U^{\rm bulk}_{\rm tot, mob}$ $\rm \left[mol/L_{bulk}\right]$, the ratio of total bulk initial uranium concentration for the immobile domain to $U^{\rm bulk}_{\rm tot,mob}$, $ratioU$ $\rm \left[-\right]$, the porosity of the immobile domain, $\theta_{\rm immob}$ $\rm \left[-\right]$, the partial pressure of CO$_2$ in the water of solution 1, $p$CO$_{2,1}$ $\rm \left[-\right]$, the pH of solution 1, $pH_1$ $\rm \left[-\right]$, the partial pressure of CO$_2$ in the water of solution 2, $p$CO$_{2,2}$ $\rm \left[-\right]$ and the pH of solution 2, $pH_2$ $\rm \left[-\right]$.

This model is quite CPU-intensive as a single run parallelized over 4 cores now usually takes about 1h30 but may rise up to 5 days for some parameter combinations. Overall, performing 500 model runs incurs a computational time of 17 days on the used workstation. Just as for problem 1, among the many simulated processes, the 1224-dimensional times series of dissolved U(VI) concentration (mol/l) in the outflow node of the mobile domain is considered as the output of interest.

\begin{table}[H]
	\caption{Uniform ranges of the free parameters considered for problems 1 and 2. The $\textbf{x}_{\rm inv}^{\rm true}$ variable is the parameter set used to generate the ``true'' data for the synthetic inverse problem 1.}
	\begin{center}
		\begin{tabular}{lcccc}%
			\hline
			Parameter & Units & Bounds & Problem & $\textbf{x}_{\rm inv}^{\rm true}$\\
			\hline
			$\mu_r$ & $\left[\rm log\left(h^{-1}\right)\right]$ & $\left[-10.5, -8.5\right]$ & 1 and 2  & -9.5\\
			$\sigma_r$ & $\left[\rm log\left(h^{-1}\right)\right]$ & $\left[2, 3\right]$ & 1 and 2  & 2.5\\
			$logK1$ & $\left[-\right]$ & $\left[-5, -3\right]$ & 1 and 2 & -4\\
			$logK2$ & $\left[-\right]$ & $\left[15.5, 17.5\right]$ & 1 and 2 & 16.5\\
			$S^{\rm bulk}_{\rm tot}$ & $\rm \left[mol/L_{bulk}\right]$ & $\left[0.005, 0.035\right]$ & 1 & 0.02 \\
			$U^{\rm bulk}_{\rm tot}$ & $\rm \left[mol/L_{bulk}\right]$ & $\left[1 \times 10^{-6}, 1 \times 10^{-5}\right]$ & 1 & 5.5 $\times$ 10$^{-6}$ \\
			$p$CO$_2$ & $\left[-\right]$ & $\left[-3.2, -2.6\right]$ & 1 & -2.9 \\
			$pH$ & $\left[-\right]$ & $\left[7.8, 8.5\right]$ & 1 & 8.15 \\
			$S^{\rm bulk}_{\rm tot,mob}$ & $\rm \left[mol/L_{bulk}\right]$ & $\left[0.005, 0.035\right]$ & 2 & \textit{NA}* \\
			$ratioS$ & $\left[-\right]$ & $\left[0.2, 5\right]$ & 2 & \textit{NA}* \\
			$U^{\rm bulk}_{\rm tot,mob}$ & $\rm \left[mol/L_{bulk}\right]$ & $\left[1 \times 10^{-6}, 1 \times 10^{-5}\right]$ & 2 & \textit{NA}* \\
			$ratioU$ & $\left[-\right]$ &  $\left[0.2, 0.5\right]$ & 2 & \textit{NA}* \\
			$\theta_{\rm immob}$ & $\left[-\right]$ & $\left[0.01, 0.1\right]$ & 2 & \textit{NA}* \\
			$p$CO$_{2,1}$ & $\left[-\right]$ & $\left[-3.2, -2.6\right]$ & 2 & \textit{NA}* \\
			$pH_1$ & $\left[-\right]$ & $\left[7.8, 8.5\right]$ & 2 & \textit{NA}* \\
			$p$CO$_{2,2}$ & $\left[-\right]$ & $\left[-2.5, -1.8\right]$ & 2 & \textit{NA}* \\
			$pH_2$ & $\left[-\right]$ & $\left[6.5, 7.5\right]$ & 2 & \textit{NA}* \\
			\hline
			\multicolumn{5}{l}{*Not applicable.} \\
		\end{tabular}
	\end{center}
	\label{table1}
\end{table}

\section{Benchmarking of the Emulators}
\label{bench}
\subsection{Used Metrics}
\label{metrics}
We resort to two metrics to assess the performance of every considered emulator for an independent test set of samples, $\left[\textbf{X}^*,\textbf{Y}^*\right]$, that was therefore not used for training. The $Q_2$ coefficient is given by
	\begin{equation}
	Q_2 = 1-\frac{\sum_{i = 1}^{n^*}\sum_{j = 1}^{d_y}\left(y^*_{i,j} - y^*_{s,i,j}\right)^2}{\sum_{i = 1}^{n^*}\sum_{j = 1}^{d_y}\left(y^*_{i,j} - \overline{\textbf{Y}^*}\right)^2},
	\label{q2}
	\end{equation}
where as stated earlier, $\textbf{Y}^*_s$ is a $n^* \times d_y$ array of simulated outputs and $\overline{\textbf{Y}^*}$ denotes the mean of $\textbf{Y}^*$. Furthermore, the root-mean-square-error (RMSE) is defined as
	\begin{equation}
	{\rm RMSE} =\sqrt{\frac{\sum_{i = 1}^{n^*}\sum_{j = 1}^{d_y}\left(y^*_{i,j} - y^*_{s,i,j}\right)^2}{n^*d_y}}.
	\label{rmse}
	\end{equation}

\subsection{Emulation Using Variable Training Set Sizes}
\label{training_set}
The 3 considered emulation methods are primarily compared for each problem in their ability to reproduce the input-output behavior of the original RTM. The ``training'' set used to construct the emulators is obtained by standard Latin hypercube sampling \citep[LHS][]{McKay1979} of the parameter hypercube defined by the bounds listed in Table \ref{table1}. The following set sizes are considered: 75, 175 and 500. Obtaining these samples by running the parallelized HPx code on our 16-cores workstation took a fair amount of time. For problem 1, about 100 samples/day can be run and collecting 500 output samples thus required 5 days. The computational demand of problem 2 is much larger and, as written already, here it took about 17 days to build the ensemble of 500 input-output examples. 

The two considered reactive transport problems both present a functional output (see section \ref{prob_description}): the 1224-dimensional time series of dissolved U(VI) concentration (mol/l) at some spatial location in the model domain. In this case it might be interesting to perform a parametric dimensionality reduction of the model output prior to the construction of an emulator. This can be done by principal component analysis (PCA), also known as proper orthogonal decomposition (POD), Karhunen-Lo\`eve (KL) transform or singular value decomposition (SVD) \citep[see, e.g., the studies by][]{Marrel2015, Roy2018}. A PCA projection can encode the linear correlations between pairs of output data points but cannot deal with more complex data dependencies. Such strategy was tested herein and it was observed that using 10 and 25 projection coefficients lead to an accurate reconstruction of the original signal \citep[with more than 95\% of the variance explained, see, e.g.,][for details about PCA]{Jolliffe2002} for problems 1 and 2, respectively. Nevertheless, using the resulting 10-dimensional and 25-dimensional outputs to build the GP and PCE emulators did not show any advantage in terms of emulation accuracy compared to working directly with the original 1224-dimensional output space directly. Regarding the used DNN method, performing a PCA of the output space prior to training is useless because, as opposed to the PCE and GP models, our DNNs try to learn all the output data dependencies simultaneously (thus not only the pairwise linear correlations but higher order statistics as well).

The GP, PCE and DNN methods not only have parameters but also hyperparameters that need to be appropriately set, e.g., covariance kernel and mean function equation for GPs, expansion order for PCE, and number of layers, neurons per layer, activation functions and amount of regularization for DNNs. Note that for the GP method, the same covariance kernel and kernel parameters are used for all of the 1224 GP-based predictions. Similarly, all of the 1224 PCE emulators share the same expansion order. For a given set of hyperparameters, actual training or learning was performed on the basis of the first 50, 150 or 475 samples, respectively, and the resulting $Q_2$ and RMSE between true and emulated data points was then estimated using the remaining 25 ``validation'' samples. This procedure allowed for selecting the optimal hyperparameter combinations across the tested ones: every combination of covariance kernel and mean function available in GPflow for the GP method, expansion order from 1 to 3 (problem 2) or 4 (problem 1) for PCE, and a few different network architectures and regularization strategies for the DNN method. After selection of the hyperparameters, each emulator was trained again using all available samples for each training set size: 75, 175 and 500.

The hyperparameters selected for the DNN approach are described in section \ref{methods_dnn}. As stated earlier, they were chosen based on limited trial-and-error using  problem 1, and were kept unchanged for problem 2. With respect to GPs, the linear mean function described in equation (\ref{gp8}) was always used while the optimal covariance kernels were different for problems 1 and 2: the sum of the SE and ACOS2 kernels (see section \ref{methods_gp}) for problem 1 and the SE kernel alone for problem 2. In contrast to the DNN and GP methods, the optimal hyperparameters for PCE and sPCE were found to be dependent on the training set size. Here the optimal PCE expansion orders for the 75, 175 and 500 training set sizes of problem 1 are 3, 2 and 3, respectively. For sPCE, these optimal values are 2, 4 and 3. For problem 2 and the same 3 sizes 75, 175 and 500, the derived optimal order values are 1, 1 and 2 for PCE and 1, 2 and 2 for sPCE, respectively.

Lastly, note that the comparison between the emulation methods is performed using an independent test set of 100 (problem 1) or 240 (problem 2) examples randomly drawn from the uniform parameter space. These test sets were thus not used to build the emulators.

\subsection{CPU-Intensive Tasks}
\label{cpu_tasks}
After looking at direct emulation capabilities, emulator performance for 3 critical CPU-intensive tasks for which emulation can be highly beneficial is scrutinized. These tasks are (in increasing order of computational requirements): global sensitivity analysis \citep[GSA, e.g.,][]{Saltelli2010,Iooss-Lemaitre2015}, Monte-Carlo (MC) based uncertainty propagation (UP), and probabilistic calibration using Markov chain Monte Carlo \citep[MCMC, e.g.,][]{Robert-Casella2004} sampling. For brevity, we only give below a very concise description of the considered GSA, UP and MCMC inversion procedures while the reader is referred to the listed books and papers for more information. For problem 1, the GSA, UP and MCMC results obtained from using the PCE, GP and DNN emulators are compared to those obtained from using the original RTM1. Because of the computational costs incurred by running the original RTM2, for problem 2 this is only done for GSA and UP. 

\subsubsection{Global Sensitivity Analysis}
\label{method_gsa}
The GSA technique aims at ranking the considered model parameters according to their influence on the model outputs across the whole parameter space, accounting for parameter interactions. This is typically done by variance-based GSA such as the Sobol' method \citep[][]{Sobol1993,Sobol2001}. The latter basically provides the so-called first-order, $S_i$, and total-order, $ST_i$, sensitivity indices. The $S_i$ index quantifies the effect of parameter $x_i$ alone, but averaged over possible variations in other input parameters. The $ST_i$ index measures the influences of $x_i$ accounting for its interactions with other parameters. There are many ways for calculating the $S_i$ and $ST_i$ indices. 

For problem 1, we invoked the ``Saltelli \& Jansen'''s formulas by \citet[][]{Saltelli2010} for which an open-source code can be downloaded from \url{https://ec.europa.eu/jrc/en/samo/simlab}. This approach requires $N \times \left(d_x+2\right)$ model evaluation points and is best used in conjunction with a low-discrepancy sequence for selecting these points. We set $N = 250$ thereby leading to 2500 evaluation points. These points were calculated from a standard low-discrepancy Sobol' sequence. For this analysis a single model output was considered: the maximum peak value across the 1224-dimensional time series of simulated U(VI) concentrations. Even though the required sample size to get accurate Sobol' indices depends on both model nonlinearity and dimensionality, a total of 2500 samples for a 8-dimensional nonlinear model is arguably low. Yet this is a pragmatic choice imposed by computational expense as running these 2500 samples through RTM1 takes about 25 days. While (very) accurate indices' estimation is out of reach, we believe that these 2500 points can be used to compare the original and emulated indices in terms of ranking of parameter influence.

The $S_i$ and $ST_i$ indices associated with RTM1 were thus first compared to those obtained by running the GP and DNN emulators for the same 2500 parameter sets. To provide more insights into the uncertainty associated with these estimates, 95\% uncertainty intervals were computed using standard bootstrap (that is, sampling with replacement from the same set of samples) using 1000 bootstrap replicas. Moreover, in order to get a taste of how the estimation process converges with the used number of samples, the GP and DNN-emulated $S_i$ and $ST_i$ indices and their uncertainty intervals were derived again using now $N = 5000$, which corresponds to a total of 50,000 evaluation points. Regarding PCE, the emulated $S_i$ and $ST_i$ values are readily calculated by manipulating the expansion coefficients \citep[see][for details]{Sudret2008}. This means that varying the number of samples does not make sense here, and bootstrapping cannot be applied.

With respect to problem 2, computational constraints make it infeasible to compute the $S_i$ and $ST_i$ indices associated with (the maximum concentration value simulated by) the original RTM2 using the Saltelli \& Jansen approach. Instead, we derived the $S_i$ indices only with the EASI method by \citet{Plischke2010}. This method uses fast Fourier transformations (FFT) and has the advantage that it does not require evaluation of the model at specific points but works with any point set. We therefore compared the $S_i$ indices of the RTM2 model and its GP and DNN emulators, obtained from using the available 240 test points. In addition, the PCE-based indices are again calculated directly.

As written already, sensitivity analysis does not necessarily require a large accuracy of the applied emulator. Hence, for this task only a tiny training set of 25 examples is also considered (in addition to the training set sizes 75, 175 and 500). Here learning and hyperparameter selection were performed using the first 20 and last 5 samples, respectively. Just as for the other training set sizes, before being used the emulators were trained again using all of the available samples (25 herein). A ``validation'' set of only 5 samples for chosing the hyperparameters is not robust. However, for the considered problems this leads to the same optimal hyperparameters as described in section \ref{training_set} for the GP and DNN methods. In contrast, every tested PCE order performs rather similarly. Therefore, for problem 1 an order-2 expansion was chosen to derive the $S_i$ and $ST_i$ indices from these 25 samples, while for problem 2 an order-1 expansion was selected.

\subsubsection{Uncertainty Propagation}
\label{method_up}
This application is concerned with the probability over the parameter space that the simulated maximum U(IV) concentration exceeds a pre-specified threshold. Threshold values of 2 $\times$ 10$^{-6}$ mol/l and 1.5 $\times$ 10$^{-6}$ mol/l are considered for problems 1 and 2, respectively. Moreover, the mean and standard deviation of the maximum U(IV) concentration are also looked at. This is done for all training set sizes: 75, 175, and 500, and using either 2500 (for problem 1, see further) or 240 (for problem 2) test samples. A total of 240 samples might seem low for the task at hand but as seen in section \ref{res_pbm2_up}, this is enough to spot large deviations from the original model's statistics. 

\subsubsection{Inverse Modeling}
\label{method_inv}
Here we focus on the problem of estimating the RTM1 model parameters from the inversion of measured U(IV) concentration data within a Bayesian framework. This is done for problem 1 and the corresponding emulators when constructed using 500 training samples.

A common representation of the forward problem is
	\begin{equation}
	\hat{\textbf{y}} = M\left(\textbf{x}\right) + \textbf{e},
	\label{mcmc0}
	\end{equation}
where the $d_y$-dimensional $\hat{\textbf{y}}$ vector contains the measurement data, and the noise term $\textbf{e}$ lumps all sources of errors. 

In the Bayesian paradigm, parameters in $\textbf{x}$ are viewed as random variables with a posterior pdf, $p\left(\textbf{x} | \hat{\textbf{y}} \right)$, that can be written as
	\begin{equation}
	p\left(\textbf{x} | \hat{\textbf{y}} \right) \propto p\left(\textbf{x}\right) L\left(\textbf{x} | \hat{\textbf{y}}\right),
	\label{mcmc1}
	\end{equation}
where $p\left(\textbf{x}\right)$ denotes the prior distribution of $\textbf{x}$ and $L \left(\textbf{x} | \hat{\textbf{y}}\right)$ signifies the likelihood function of $\textbf{x}$. 

To avoid numerical over- or underflow, it is convenient to work with the logarithm of $L \left(\textbf{x} | \hat{\textbf{y}}\right)$ (log-likelihood): $\ell\left(\textbf{x} | \hat{\textbf{y}}\right)$. If we assume $\textbf{e}$ to be normally distributed, uncorrelated and heteroscedastic with variances, $\sigma_{e,1}^2, \cdots, \sigma_{e,d_y}^2$, $\ell\left(\textbf{x} | \hat{\textbf{y}}\right)$ can be written as

\begin{equation}
\ell\left(\textbf{x} | \hat{\textbf{y}}\right) = -\frac{d_y}{2}\log\left(2 \pi\right) - \frac{1}{2}\log\left(\prod_{i = 1}^{d_y}\sigma_{e,i}^2\right) -\frac{1}{2}\sum_{i = 1}^{d_y} \left[ \frac{\hat{y}_i - M_i\left(\textbf{x}\right)} {\sigma_{e,i}} \right]^2,
\label{mcmc2}
\end{equation}
where the $M_i\left(\textbf{x}\right)$ are the simulated responses that are compared with the $i = 1, \cdots, d_y$ measurement data, $\hat{y}_i$.

Since we consider herein a synthetic inverse problem, we created the ``true'' measurements, $\hat{\textbf{y}}$, by running RTM1 with the $\textbf{x}_{\rm inv}^{\rm true}$ parameter set listed in Table \ref{table1} and corrupting the resulting 1224-dimensional output with a zero-mean Gaussian white noise with standard deviations fixed to 3\% of the simulated values. A relative noise of 3\% seems typical for U(VI) concentration data \citep[][]{Yin2011}.

An exact analytical solution of $p \left(\textbf{x} | \hat{\textbf{y}}\right)$ is not available for the type of non-linear inverse problems considered herein. We therefore resort to MCMC simulation \citep[see, e.g.,][]{Robert-Casella2004}. More specifically, the DREAM$_{\rm \left(ZS\right)}$ algorithm is used to approximate the posterior distribution. A detailed description of this sampling scheme including a proof of ergodicity and detailed balance can be found in \citet{Vrugt2009} and \citet{Laloy-Vrugt2012}. 

\section{Results}
\label{results}

\subsection{Problem 1}
\label{res_pbm1}

\subsubsection{Emulation}
\label{res_pbm1_emul}

\begin{figure}[H]
	\noindent\hspace{0cm}\includegraphics[width=35pc]{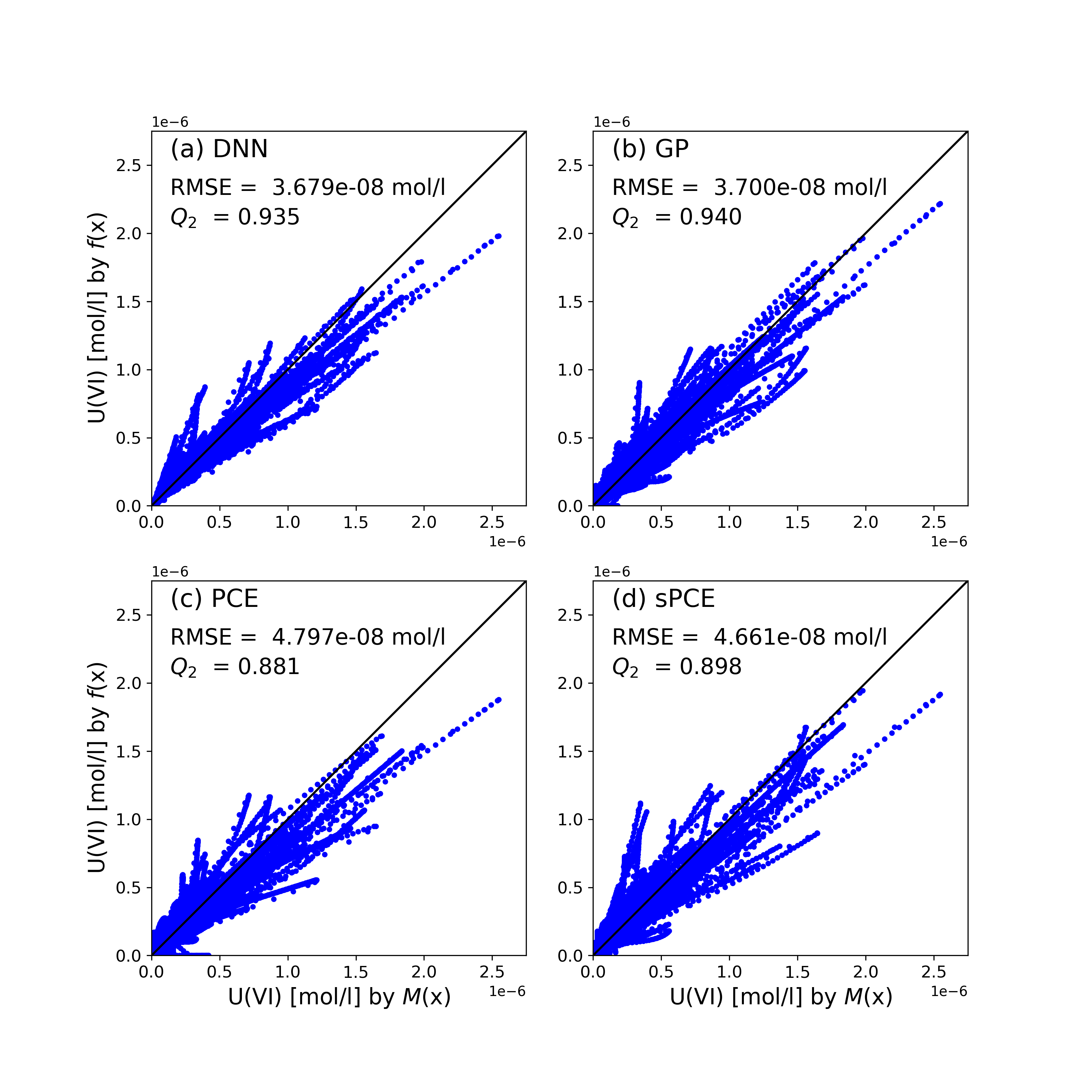}
	\caption{1-1 plots of emulation performance obtained for problem 1 when the emulators are built using 75 samples. The $x$-axis and $y$-axis present the true and emulated 100 $\times$ 1224 test data points, respectively. The RMSE and $Q_2$ coefficient denote the root-mean-square-error and coefficient of determination in testing mode, respectively, between the true and emulated 100 $\times$ 1224 test data points.}
	\label{fig1}
\end{figure}

\begin{figure}[H]
	\noindent\hspace{0cm}\includegraphics[width=35pc]{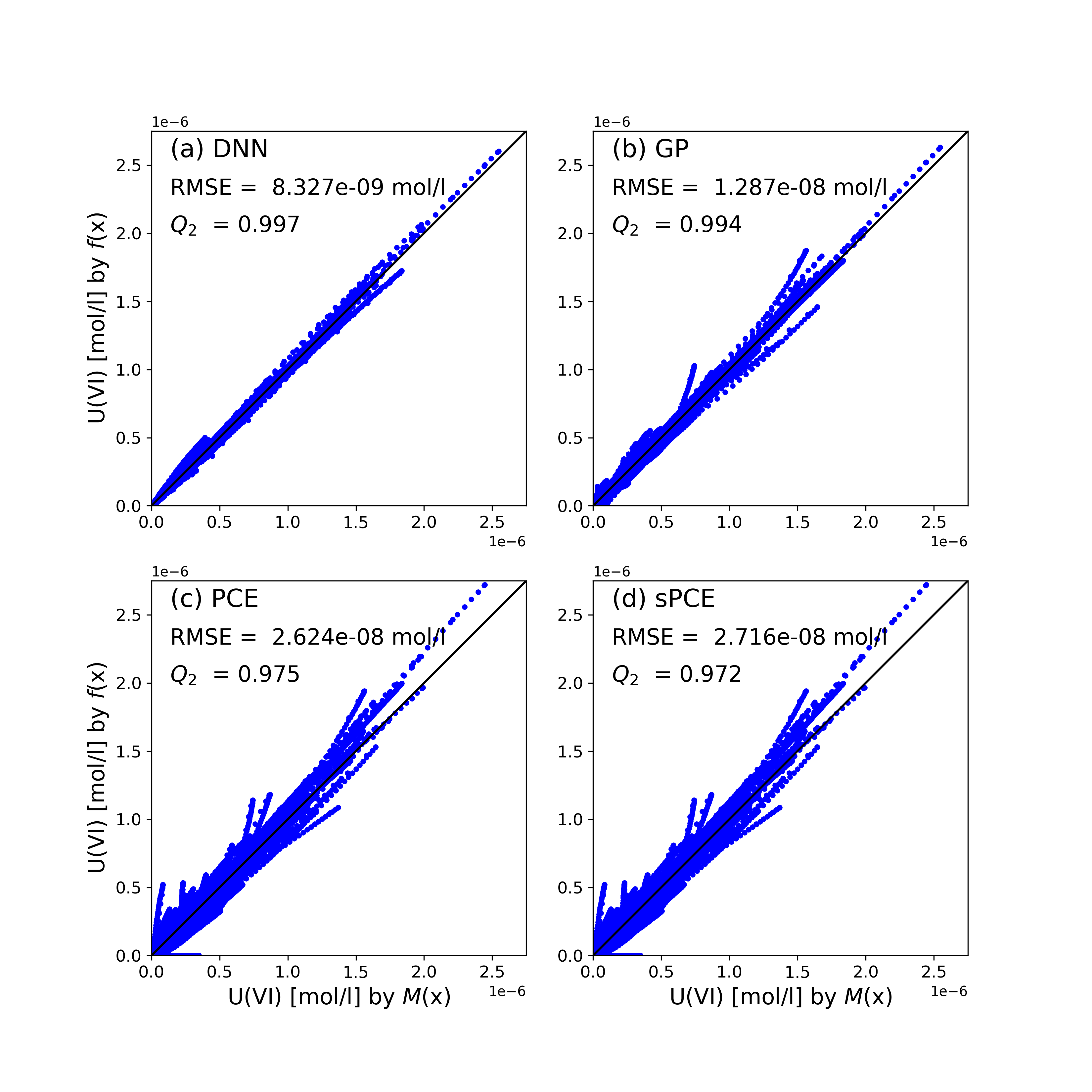}
	\caption{1-1 plots of emulation performance obtained for problem 1 when the emulators are built using 500 samples. The $x$-axis and $y$-axis present the true and emulated 100 $\times$ 1224 test data points, respectively. The RMSE and $Q_2$ coefficient denote the root-mean-square-error and coefficient of determination in testing mode, respectively, between the true and emulated 100 $\times$ 1224 test data points.}
	\label{fig2}
\end{figure}

Figures \ref{fig1}-\ref{fig2} depict the emulation results obtained when using a total of 75 and 500 samples for constructing the emulators. Table \ref{table2} presents the same results for the case where a set of 175 training samples is used. As mentioned in section \ref{training_set}, these results are for a test set of 100 samples randomly drawn from the uniform 8-dimensional parameter space. It is seen that the DNN and GP emulators always outperform the PCE and sPCE emulators, with a better performance of the DNNs compared to the GPs. Also, the two worst performing emulators, PCE and sPCE, do not differ much in accuracy. Therefore, the sPCE emulators will no longer be considered in the rest of this section dedicated to problem 1 (section \ref{res_pbm1}).

\begin{table}[H]
	\caption{Emulation performance obtained for the 100-sample test set of problem 1 when the emulators are built using 175 samples. The RMSE and $Q_2$ coefficient denote the root-mean-square-error and coefficient of determination in testing mode, respectively, between the true and emulated 100 $\times$ 1224 test data points.}
	\begin{center}
		\begin{tabular}{lcc}%
			\hline
			Emulator & $Q_2$ & RMSE [mol/l]\\
			\hline
			DNN & 0.994 & 1.235 $\times$ 10$^{-8}$ \\
			GP & 0.979 & 2.482 $\times$ 10$^{-8}$  \\
			PCE & 0.953 & 3.590 $\times$ 10$^{-8}$  \\
			sPCE & 0.914 & 4.978 $\times$ 10$^{-8}$  \\
			\hline
		\end{tabular}
	\end{center}
	\label{table2}
\end{table}

\begin{figure}[H]
	\noindent\hspace{0cm}\includegraphics[width=40pc]{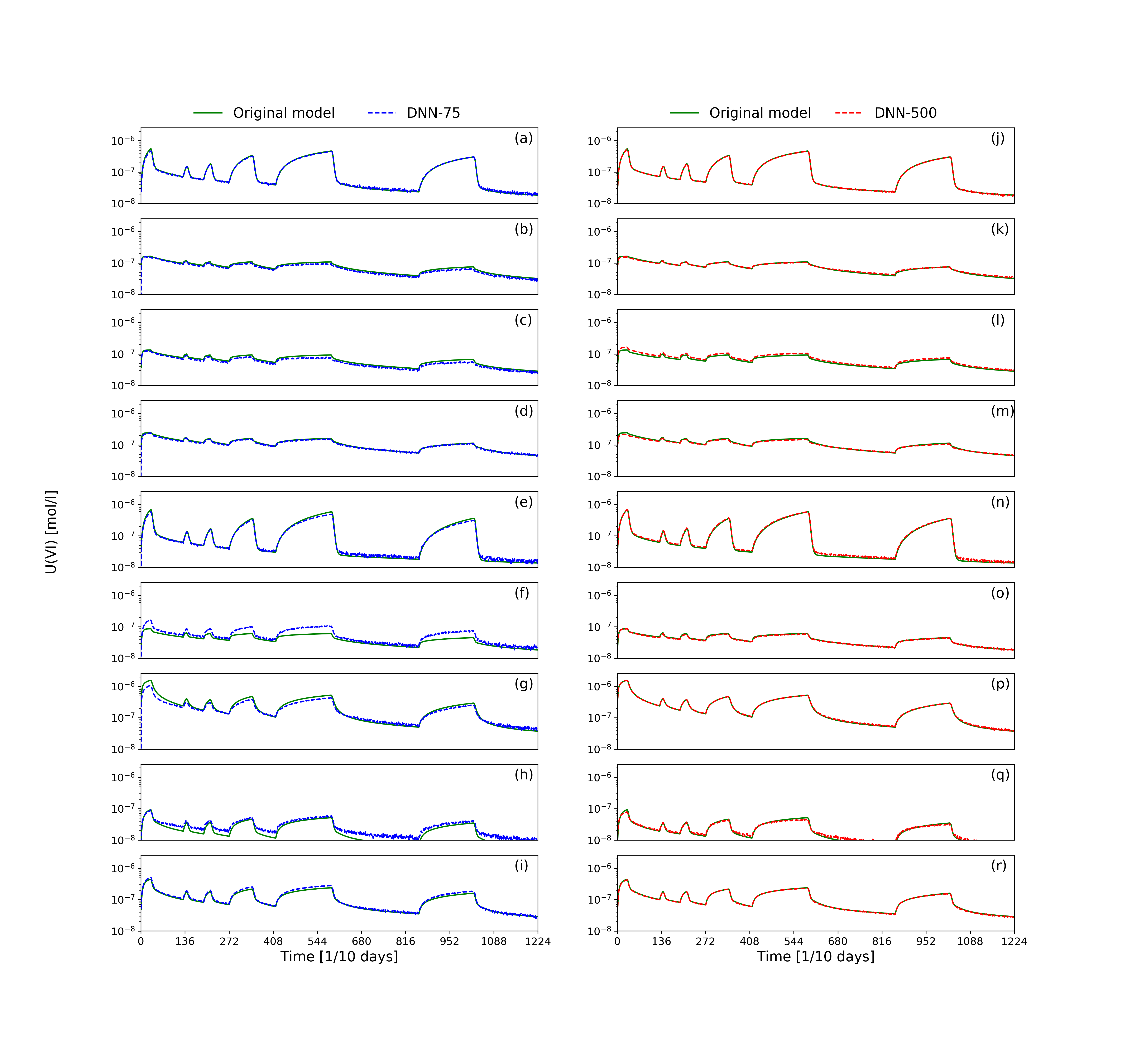}
	\caption{Original (green solid lines) and DNN-emulated (blue and red dashed lines) time series of 1224 U(VI) concentrations for problem 1. Each of the 9 rows corresponds to a different test example randomly chosen from the ensemble of 100 test examples. DNN-75 (blue dashed lines) signifies a DNN built using 75 training examples. DNN-500 (red dashed lines) denotes a DNN constructed using 500 training examples.}
	\label{fig3}
\end{figure}

\begin{figure}[H]
	\noindent\hspace{0cm}\includegraphics[width=40pc]{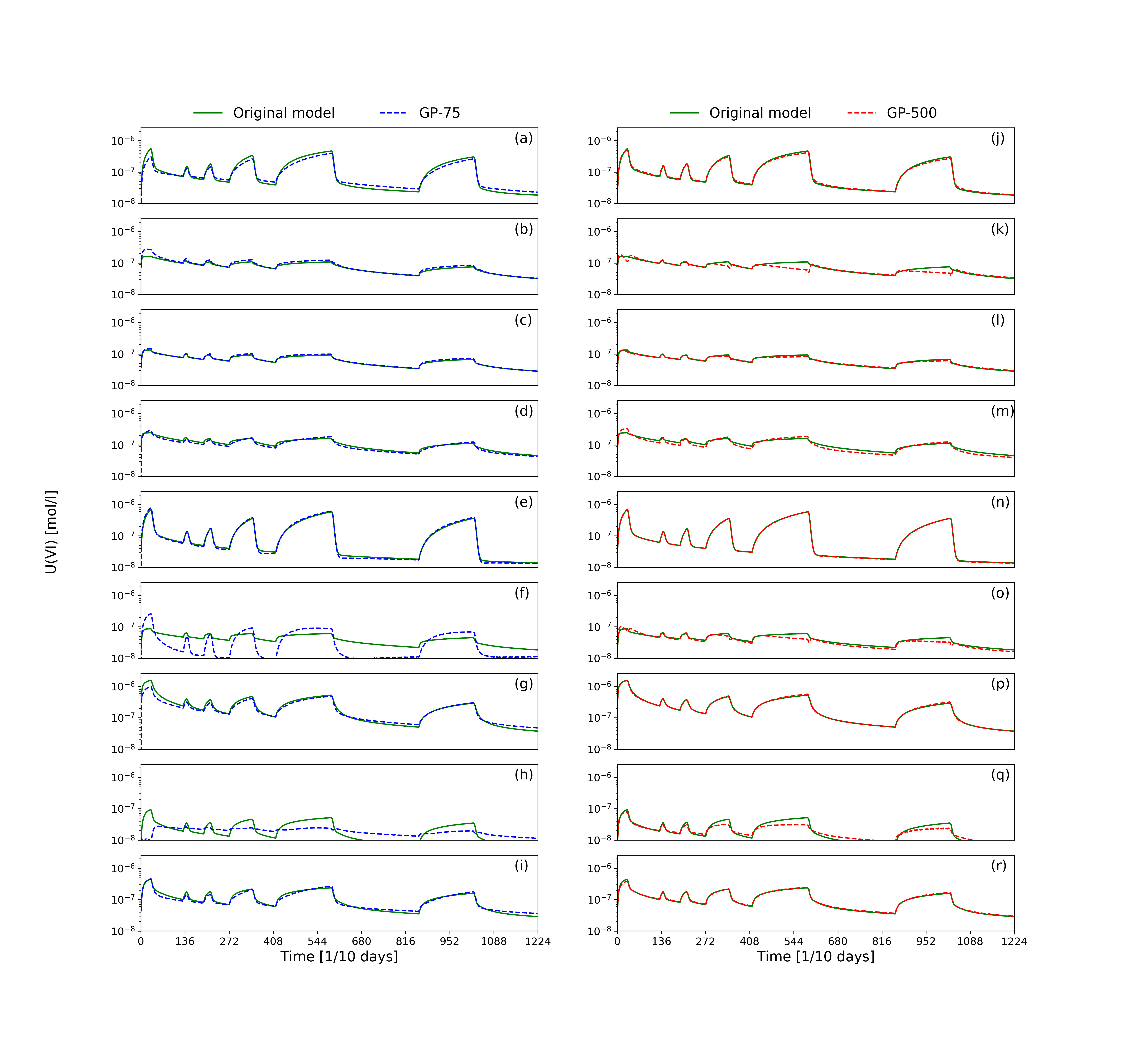}
	\caption{Original (green solid lines) and GP-emulated (blue and red dashed lines) time series of 1224 U(VI) concentrations for problem 1. Each of the 9 rows corresponds to a different test example randomly chosen from the ensemble of 100 test examples. GP-75 (blue dashed lines) signifies a GP built using 75 training examples. GP-500 (red dashed lines) denotes a GP model constructed using 500 training examples.}
	\label{fig4}
\end{figure}

Figures \ref{fig3}-\ref{fig4} provide more insights into the emulation accuracy of the DNN and GP methods. These figures depict the original and emulated time series of U(VI) concentrations for 9 examples randomly chosen from the test set, for the cases where the emulator is constructed using 75 (DNN-75, GP-75) or 500 (DNN-500, GP-500) training examples. Strikingly, the DNN-500 emulator best mimics the original RTM with emulated curves that are visually extremely close to their original counterparts (Figures \ref{fig3}j-r). Nevertheless, the DNN-emulated curves show a slight but seemingly complex noise. Although the GP-emulated curves clearly do no reproduce the original ones a accurately as the DNN-emulated curves, they are much smoother.

\subsubsection{Sensitivity Analysis}
\label{res_pbm1_gsa}
\begin{figure}[H]
	\noindent\hspace{0cm}\includegraphics[width=35pc]{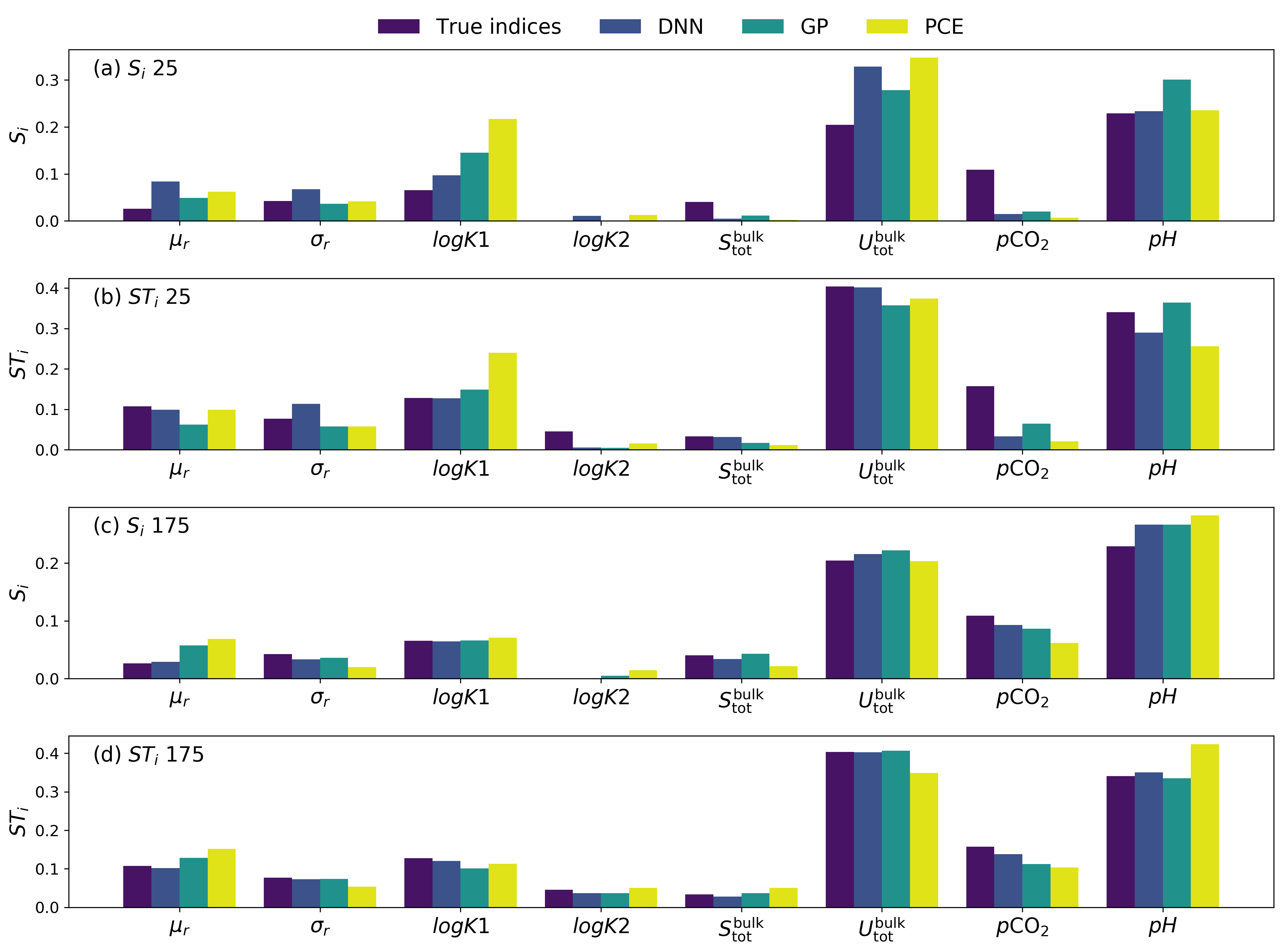}
	\caption{First ($S_i$) and total ($ST_i$) Sobol' sensitivty indices for the peak concentration associated with problem 1. $S_i$ 25 and $ST_i$ 25 mean the emulated indices derived from using 25 training samples to contruct the emulators. Similarly, $S_i$ 175 and $ST_i$ 175 mean the emulated indices derived from using 175 training samples to contruct the emulators. The true indices are derived using 2500 points from a Sobol' low-discrepancy sequence. The DNN and GP-based indices are derived using the same 2500 points. As commony done with PCE, the PCE-based indices are obtained directly from manipulating the expansion coefficients. Yet feeding the PCE emulators with the 2500 points to calculate the indices leads to very similar results (not shown).}
	\label{fig5}
\end{figure}

Figure \ref{fig5} displays the ``true'' and emulated $S_i$ and $ST_i$ indices associated with the maximum (peak) U(VI) concentration. The emulated indices are derived from using 25 and 175 training examples (see section \ref{method_gsa} for details). Overall, the DNN-175, GP-175 and PCE-175 methods appear to perform similarly and relatively well. As expected, the approximation of the $S_i$ and $ST_i$ indices worsen as the training set size decreases to 25. For this rather small training set, the DNN-25 and GP-25 emulators offer a slightly better performance than the PCE-25 emulator though. In addition, although far from perfect the indices derived by all 3 methods when constructed with the training size 25 nevertheless correctly identify the 2 most influential parameters.

Table \ref{table3} presents the 95\% uncertainty intervals associated with the indices shown in Figure \ref{fig5} (for the Saltelli \& Jansen's approach only), together with the $S_i$ indices derived by the EASI method using the same 2500 samples. For comparative purposes, the emulated $S_i$ and $ST_i$ indices produced by the DNN and GP methods when setting $N = 5000$ in the Saltelli \& Jansen's approach, thereby leading to a total of 50,000 samples, are also presented. It is seen that for the original (RTM1) model, the Saltelli \& Jansen's and EASI approaches produce the same parameter ranking and globally similar $S_i$ values  (compare 1$^{\rm st}$ and 9$^{\rm th}$ rows of Table \ref{table3}). The largest difference is observed for the most influential $pH$ parameter for which the Saltelli \& Jansen's formulas give $S_i = 0.22$ while EASI gives $S_i = 0.28$. Furthermore, moving from using a total of 2500 samples to a total 50,000 samples with the Saltelli \& Jansen's approach does not change the $S_i$ ranking produced by the DNN-175 and GP-175 emulators (compare 2$^{\rm nd}$ and 3$^{\rm rd}$ rows of Table \ref{table3} with 12$^{\rm th}$ and 13$^{\rm th}$ rows of Table \ref{table3}, respectively). However, the $ST_i$ ranking derived by DNN-175 and GP-175 varies: here the two most influential parameters, $U^{\rm bulk}_{\rm tot}$ and $pH$, switch ranks (compare 4$^{\rm th}$ and 5$^{\rm th}$ rows of Table \ref{table3} with 14$^{\rm th}$ and 15$^{\rm th}$ rows of Table \ref{table3}, respectively). The uncertainty intervals associated with the indices also become substantially narrower (for instance, compare 2$^{\rm nd}$ and 12$^{\rm th}$ rows of Table \ref{table3}).

\begin{table}[H]
	\scriptsize
	\caption{\small{$S_i$ and $ST_i$ sensitivity indices associated with the original model (RTM1) and the emulators constructed with 175 training samples (DNN-175, GP-175, and PCE-175), for different estimation methodologies. ``Saltelli \& Jansen - 2500'' and ``Saltelli \& Jansen - 50,000'' signify the Saltelli \& Jansen's approach with a total of 2500 and 50,000 samples, respectively. ``EASI - 2500'' means the EASI approach with 2500 samples. The values between brackets denote the lower and upper bounds of the 95\% uncertainty intervals derived using bootstrap (Saltelli \& Jansen's approach only).}}
	\begin{center}
		\begin{adjustbox}{angle=90}
			\hspace*{-2cm}\begin{tabular}{lcccccccc}%
				\hline
				Index & $\mu_r$ & $\sigma_r$ & $logK1$ & $logK2$ & $S^{\rm bulk}_{\rm tot}$ & $U^{\rm bulk}_{\rm tot}$ & $p$CO$_2$ & $pH$\\
				\hline
				\multicolumn{8}{c}{ }\\
				\multicolumn{8}{c}{Saltelli \& Jansen - 2500}\\
				\multicolumn{8}{c}{ }\\
				$S_i$ RTM1 & 0.03 $\left[0,0.10\right]$ & 0.04 $\left[0,0.11\right]$ & 0.07 $\left[0,0.15\right]$ & 0.00 $\left[0.00,0.04\right]$ & 0.04 $\left[0,0.09\right]$& 0.20 $\left[0.05, 0.38\right]$ & 0.11 $\left[0.10 ,0.21\right]$ & 0.22 $\left[0.09,0.38\right]$\\ 
				$S_i$ DNN-175 & 0.03 $\left[0,0.09\right]$ & 0.03 $\left[0,0.11\right]$ & 0.06 $\left[0,0.15\right]$ & 0.00 $\left[0.00,0.04\right]$& 0.03 $\left[0.00,0.09\right]$& 0.21 $\left[0.06,0.38\right]$& 0.09 $\left[0.00,0.19\right]$& 0.27 $\left[0.13,0.42\right]$\\
				$S_i$ GP-175 & 0.06 $\left[0,0.13\right]$& 0.04 $\left[0,0.12\right]$& 0.07 $\left[0,0.15\right]$& 0.00 $\left[0.00,0.05\right]$& 0.04 $\left[0.00,0.11\right]$& 0.22 $\left[0.07,0.39\right]$& 0.09 $\left[0,0.18\right]$& 0.27$\left[0.12,0.43\right]$\\
				$ST_i$ RTM1 & 0.11 $\left[0.07,0.16\right]$ & 0.08 $\left[0.04,0.12\right]$ & 0.13 $\left[0.08,0.20\right]$ & 0.05 $\left[0.02,0.08\right]$ & 0.03 $\left[0.02,0.05\right]$ & 0.40 $\left[0.28,0.56\right]$& 0.16 $\left[0.10,0.24\right]$& 0.34 $\left[0.26,0.44\right]$\\
				$ST_i$ DNN-175 & 0.10 $\left[0.07,0.15\right]$& 0.07 $\left[0.04,0.11\right]$& 0.12 $\left[0.07,0.18\right]$& 0.04 $\left[0.02,0.06\right]$& 0.03 $\left[0.02,0.04\right]$& 0.40 $\left[0.29,0.54\right]$& 0.14 $\left[0.09,0.20\right]$& 0.35 $\left[0.27,0.46\right]$\\
				$ST_i$ GP-175 & 0.13 $\left[0.09,0.18\right]$& 0.07 $\left[0.04,0.11\right]$& 0.10 $\left[0.07,0.14\right]$& 0.04 $\left[0.03,0.05\right]$& 0.04 $\left[0.03,0.05\right]$& 0.41 $\left[0.28,0.55\right]$& 0.11 $\left[0.08,0.15\right]$& 0.34 $\left[0.26,0.43\right]$\\
				\multicolumn{8}{c}{ }\\
				\multicolumn{8}{c}{PCE}\\
				\multicolumn{8}{c}{ }\\
				$S_i$ PCE-175 & 0.07 & 0.02 & 0.07 & 0.01 & 0.02 & 0.20 & 0.06 & 0.28\\
				$ST_i$ PCE-175 & 0.15 & 0.05 & 0.11 & 0.05 & 0.05 & 0.35 & 0.10 & 0.42\\
				\multicolumn{8}{c}{ }\\
				\multicolumn{8}{c}{EASI - 2500}\\
				\multicolumn{8}{c}{ }\\
				$S_i$ RTM1 & 0.05 & 0.04 & 0.07 & 0.01 & 0.03 & 0.21 & 0.07 & 0.28 \\
				$S_i$ DNN-175 & 0.05 & 0.04 & 0.07 & 0.01 & 0.03 & 0.23 & 0.06 & 0.29 \\
				$S_i$ GP-175 & 0.06 & 0.03 & 0.07 & 0.01 & 0.04 & 0.22 & 0.06 & 0.29\\
				\multicolumn{8}{c}{ }\\
				\multicolumn{8}{c}{Saltelli \& Jansen - 50,000}\\
				\multicolumn{8}{c}{ }\\
				$S_i$ DNN-175 & 0.06 $\left[0.04,0.08\right]$& 0.02 $\left[0.01,0.04\right]$& 0.06 $\left[0.05,0.09\right]$& 0.01 $\left[0,0.02\right]$& 0.01 $\left[0,0.02\right]$& 0.22 $\left[0.20,0.25\right]$& 0.07 $\left[0.04,0.09\right]$& 0.28 $\left[0.25,0.34\right]$\\
				$S_i$ GP-175  & 0.07 $\left[0.05,0.09\right]$& 0.02 $\left[0.01,0.03\right]$& 0.06 $\left[0.05,0.08\right]$& 0.02 $\left[0,0.03\right]$& 0.02 $\left[0.01,0.03\right]$& 0.22 $\left[0.19,0.26\right]$& 0.06 $\left[0.05,0.09\right]$& 0.28 $\left[0.23,0.32\right]$\\
				$ST_i$ DNN-175 & 0.14 $\left[0.13,0.15\right]$& 0.06 $\left[0.05,0.07\right]$& 0.12 $\left[0.11,0.13\right]$& 0.03 $\left[0.03,0.04\right]$& 0.03 $\left[0.03,0.03\right]$& 0.36 $\left[0.33,0.38\right]$& 0.12 $\left[0.11,0.13\right]$& 0.42 $\left[0.39,0.45\right]$\\
				$ST_i$ GP-175 & 0.15 $\left[0.14,0.17\right]$& 0.06 $\left[0.05,0.06\right]$& 0.10 $\left[0.09,0.11\right]$& 0.03 $\left[0.03,0.04\right]$& 0.03 $\left[0.03,0.03\right]$& 0.36 $\left[0.34,0.39\right]$& 0.10 $\left[0.09,0.11\right]$& 0.41 $\left[0.39,0.44\right]$\\
				\hline
			\end{tabular}
		\end{adjustbox}
	\end{center}
	\label{table3}
\end{table}

\subsubsection{Uncertainty Propagation}
\label{res_pbm1_up}

Table \ref{table4} lists the true and emulated: (i) probability of exceeding the maximum concentration of 2 $\times$ 10$^{-6}$ mol/l, $p_{\rm MAX}$, (ii) mean peak concentration, $\mu_{\rm MAX}$, and (iii) standard deviation of the maximum concentration, $\sigma_{\rm MAX}$. The true and emulated values were all calculated over the 2500 Sobol' samples used in section \ref{res_pbm1_gsa}. It appears that all emulators built using 75 training samples perform rather badly. However, for training set sizes 175 and 500, the DNN emulators provide the best approximations to the true statistics. Then come the GP emulators before the worst performing PCE emulators.

\begin{table}[H]
	\caption{Uncertainty propagation results obtained for problem 1 from using the 3 emulators, DNN, GP and PCE, and 3 training set sizes for emulator's construction: 75, 175 and 500.}
	\small{
		\begin{center}
			\begin{tabular}{ccccccccc}%
				\hline
				DNN-75 & GP-75 & PCE-75 & DNN-175 & GP-175 & PCE-175 & DNN-500 & GP-500 & PCE-500 \\
				\multicolumn{9}{c}{$\large{{p}_{\rm MAX}}$, original model value is 0.0348}\\
				0.0188 & 0.0172 & 0.0156 & 0.0340 & 0.0320 & 0.0256 & 0.0352 & 0.0328 & 0.0308 \\
				\multicolumn{9}{c}{ }\\
				\multicolumn{9}{c}{$\large{\mu_{\rm MAX}}$ (10$^{-7}$ mol/l), original model is 5.863}\\
				5.586 & 5.941 & 5.739 & 5.806 & 6.116 & 6.246 & 5.864 & 5.916 & 5.996 \\
				\multicolumn{9}{c}{ }\\
				\multicolumn{9}{c}{$\large{\sigma_{\rm MAX}}$ (10$^{-7}$ mol/l), original model is 5.907}\\
				5.037 & 5.088 & 4.873 & 5.708 & 5.607 & 5.472 & 5.836 & 5.714 & 5.593 \\
				\hline
			\end{tabular}
		\end{center}
	}
	\label{table4}
\end{table}

\subsubsection{Inverse Modeling}
\label{res_pbm1_inv}

\begin{figure}[H]
	\noindent\hspace{0cm}\includegraphics[width=40pc]{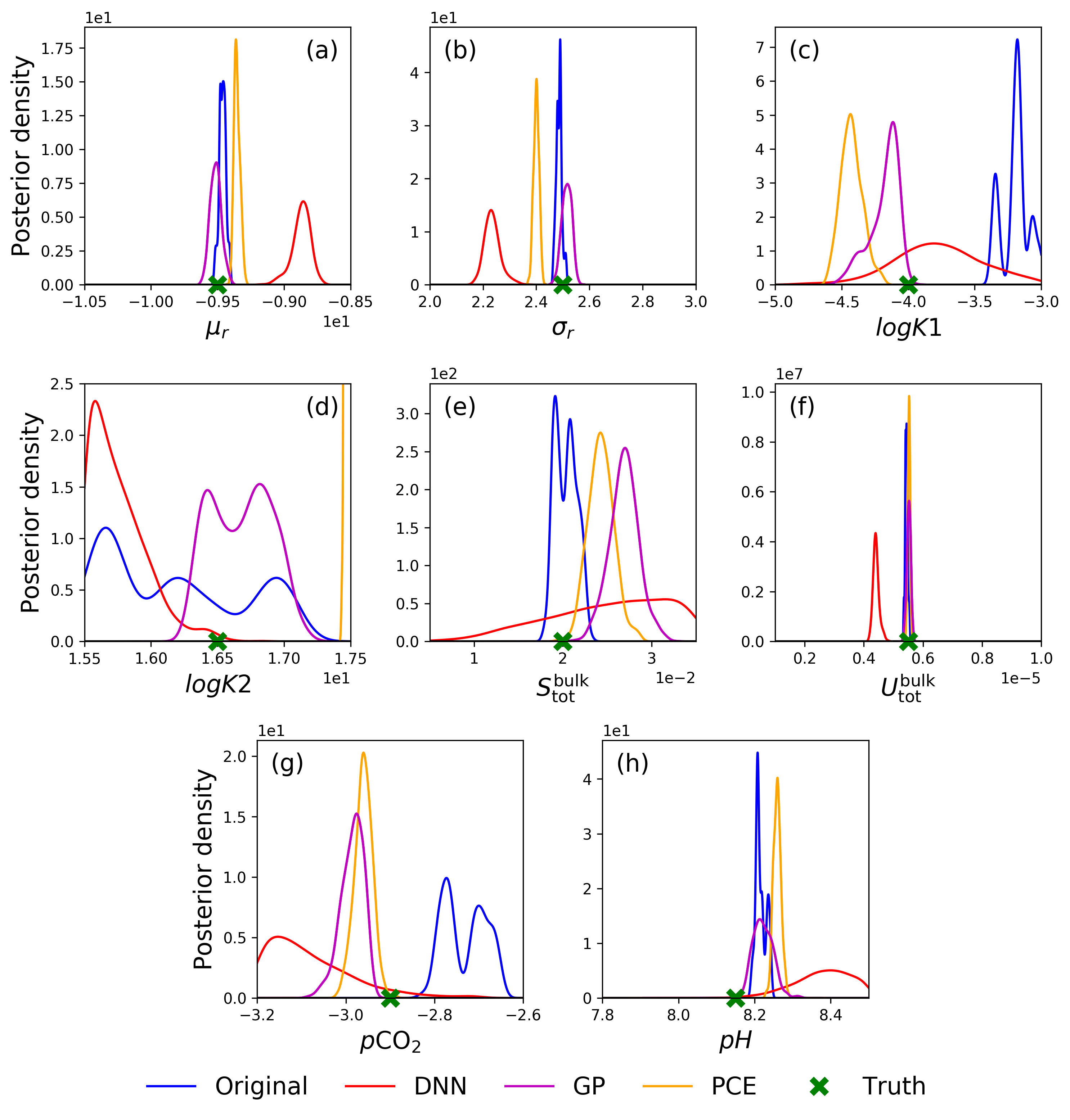}
	\caption{Marginal posterior distributions of the 8 parameters of RTM1. The blue lines (labeled as Original) denote the distributions obtained when sampling the original RTM1 model with DREAM$_{\rm \left(ZS\right)}$ for 10,000 MCMC iterations and discarding the first 7500 samples (see main text). The red, magenta, and orange lines represent the distributions obtained when sampling the posterior pdf of the DNN, GP and PCE parameters, respectively, with DREAM$_{\rm \left(ZS\right)}$ for 300,000 MCMC iterations and discarding all the samples before official convergence by the \citet{Gelman-Rubin1992} diagnostic is declared. Every set of posterior samples is smoothed by kernel density smoothing before plotting. The green crosses are the true model parameter values.}
	\label{fig6}
\end{figure}

To establish a reference case, the original RTM1 was sampled with DREAM$_{\rm \left(ZS\right)}$ for 10,000 MCMC iterations. This took about 100 days on the used workstation. The sampled log-likelihood values (see equation \ref{mcmc2}) reach equilibrium around the ``true'' log-likelihood of about 22398.4 after some 7,500 MCMC iterations (not shown). This means that from iteration 7500 on, the MCMC sampling returns posterior samples.  However, after the computational budget of 10,000 forward model runs has been consumed the sampling process is still far from having appropriately explored the posterior distribution. This is indicated by the facts that (1) the posterior distribution sampled so far (which consists of the last 2500 MCMC samples) does not encapsulate every dimension of the true parameter set (see Figure \ref{fig6}) and (2) the \citet{Gelman-Rubin1992} convergence diagnostic, $\hat{R}$, is still not satisfied (i.e., $\hat{R} \leq 1.2$) for any of the 8 sampled parameters \citep[see, e.g.,][for details about the use of $\hat{R}$ with DREAM$_{\rm \left(ZS\right)}$]{Laloy2015}. At this stage the MCMC has thus only explored a local mode of the posterior target. Such difficulty to converge within the allowed 10,000 iterations is likely caused by the low heteroscedastic noise (3\%, see section \ref{method_inv}) used to create the ``true'' measurements, which together with the relatively large number of data (1224) induces a quite peaky log-likelihood.

The DNN, GP and PCE emulators were used to replace the original RTM1 within the MCMC sampling, each time for a total of 300,000 MCMC iterations. This took less than 5 minutes for the DNN emulator, about 1 hour for the GP emulator and some 24 hours for the PCE emulator. The much larger CPU-cost incurred by the PCE method is caused by the need to evaluate a separated PCE to predict each of the 1224 components of $\textbf{y}_s$. For all 3 emulators, official $\hat{R}$-convergence is a achieved within 100,000 to 200,000 MCMC iterations. Quite surprisingly, even though the DNN emulator far surpasses the GP and PCE emulators with respect to direct emulation (see section \ref{res_pbm1_emul}, Figures \ref{fig1}-\ref{fig4} and Table \ref{table2}), when used for model calibration it leads to the largest bias in the inferred parameters. Moreover, this bias in the derived posterior distribution is much more significant than that observed for the GP and PCE emulators (see Figure \ref{fig6}). Indeed, the mode of the DNN-derived posterior distribution is often well remote from the true parameter values while for 4 parameters out of 8 the DNN-based distribution does not even include the true value (Figure \ref{fig6}).

The GP emulator offers the most accurate estimates to the true parameter values (Figure \ref{fig6} and Table \ref{table5}). The associated 95\% posterior uncertainty intervals either include or are close to the true values, while being simultaneously narrow with respect to the uniform prior distribution (Table \ref{table4}). This is consistent with the observation that among the 3 emulators, only the used GP allows for sampling appropriate log-likelihood values, that is, values that are close to 22,398. On the contrary, MCMC with PCE results in equilibrium log-likelihood values in the range 22,353-22,358, while for DNN-based MCMC this range decreases to about 22,232-22,235. This means that the PCE-based and DNN-based solutions underfit the data. This underfitting can be considered as rather slight since it is almost not noticeable visually: for each trial (original RTM1, DNN, GP, PCE), the derived maximum a posteriori (MAP) solutions look visually equally good (not shown). Yet the underfitting is large enough to strongly bias the corresponding inversion results.

The surprisingly bad performance of the DNN emulator can be at least partially attributed to the small but complex deterministic noise that affect the DNN-based predictions (see section \ref{res_pbm1_emul} and Figures \ref{fig3}-\ref{fig4}). Indeed, smoothing the time series outputted by the DNN is found to permit finding somewhat better posterior solutions by the MCMC (not shown). Nevertheless, limited testing with various amounts of smoothing combined with different smoothing methods (e.g., median filtering, Savitzky-Golay filter) did not correct all the bias and overall, did not bring the DNN results to the quality of those associated with the used GP.

Lastly, we would like to stress that in the above MCMC sampling procedure we did not attempt to correct for the emulator's bias. This complicated task is a subfield of research on its own and is beyond the scope of the current study. To account for emulation errors, we just tried to inflate the $\bm{\upsigma}_e$ vector in equation (\ref{mcmc2}) by multiplying it by a prescribed constant factor and this simple strategy did not prove very successful (not shown).

\begin{table}[H]
	\caption{95\% posterior uncertainty intervals derived from the MCMC inversion of the 1224 synthetic data associated with problem 1. The term ``original'' refers to the MCMC trial that uses the original RTM1 model.}
	\small{
		\begin{center}
			\begin{tabular}{ccccccc}%
				\hline
				Parameter & True & Prior & 95\% Original & 95\% DNN & 95\% GP & 95\% PCE \\
				\hline
				$\mu_r$ &  -9.5 & $\left[-10.5, -8.5\right]$ & $\left[-9.52, -9.41\right]$ & $\left[-9.05, -8.75\right]$ & $\left[-9.59, -9.42\right]$ & $\left[-9.39, -9.30\right]$\\
				$\sigma_r$ & 2.5 & $\left[2, 3\right]$ & $\left[2.47, 2.51\right]$ & $\left[2.18, 2.31\right]$ & $\left[2.48, 2.55\right]$ & $\left[2.38, 2.42\right]$ \\
				$logK1$ & -4 & $\left[-5, -3\right]$ & $\left[-3.36, -3.01\right]$ & $\left[-4.45, -3.11\right]$ & $\left[-4.45, -4.03\right]$ & $\left[-4.57, -4.23\right]$\\
				$logK2$ & 16.5 & $\left[15.5, 17.5\right]$ & $\left[15.55, 17.04\right]$ & $\left[15.51, 16.32\right]$ & $\left[16.32, 17.07\right]$ & $\left[17.45, 17.50\right]$ \\
				$S^{\rm bulk}_{\rm tot}$ & 0.02 & $\left[0.005, 0.035\right]$ & $\left[0.019, 0.023\right]$ & $\left[0.011, 0.035\right]$ & $\left[0.023, 0.03\right]$ & $\left[0.022, 0.027\right]$\\
				$U^{\rm bulk}_{\rm tot}$ ($\times$ 10$^{-6}$) & 5.5 & $\left[1, 10\right]$ & $\left[5.35, 5.54\right]$ & $\left[4.23, 4.67\right]$ & $\left[5.38, 5.65\right]$ & $\left[5.44, 5.60\right]$\\
				$p$CO$_2$ & -2.9 & $\left[-3.2, -2.6\right]$ & $\left[-2.80, -2.66\right]$ & $\left[-3.20, -2.83\right]$ & $\left[-3.05, -2.95\right]$ & $\left[-3.00, -2.92\right]$ \\
				$pH$ & 8.15 & $\left[7.8, 8.5\right]$ & $\left[8.19, 8.24\right]$ & $\left[8.21, 8.49\right]$& $\left[8.18, 8.28\right]$ & $\left[8.24, 8.28\right]$\\
				\hline
			\end{tabular}
		\end{center}
	}
	\label{table5}
\end{table}

\subsection{Problem 2}
\label{res_pbm2}

\subsubsection{Emulation}
\label{res_pbm2_emul}

\begin{figure}[H]
	\noindent\hspace{0cm}\includegraphics[width=35pc]{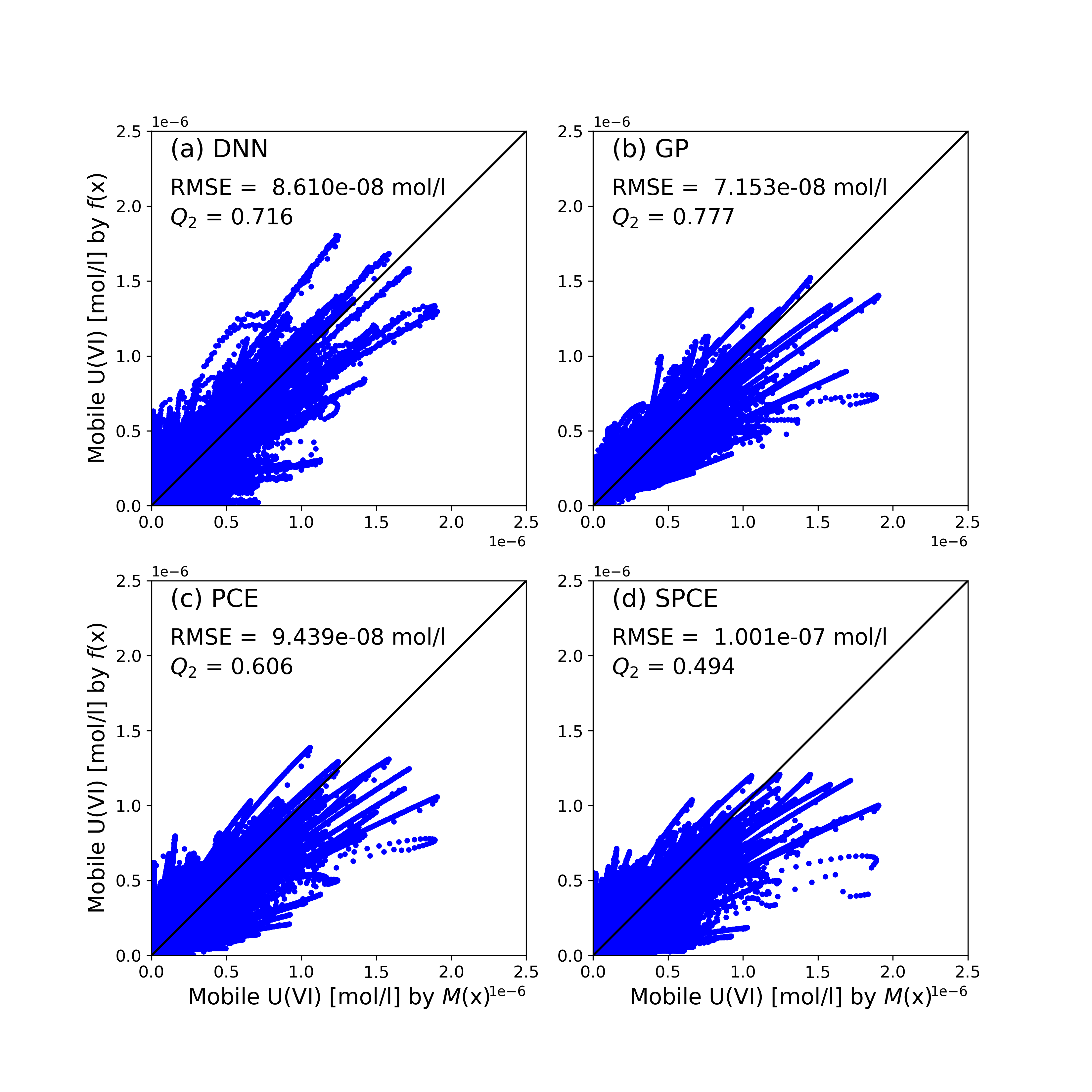}
	\caption{1-1 plots of emulation performance obtained for problem 2 when the emulators are built using 75 samples. The $x$-axis and $y$-axis present the true and emulated 240 $\times$ 1224 test data points, respectively. The RMSE and $Q_2$ coefficient denote the root-mean-square-error and coefficient of determination in testing mode, respectively, between the true and emulated 240 $\times$ 1224 test data points.}
	\label{fig7}
\end{figure}

\begin{figure}[H]
	\noindent\hspace{0cm}\includegraphics[width=35pc]{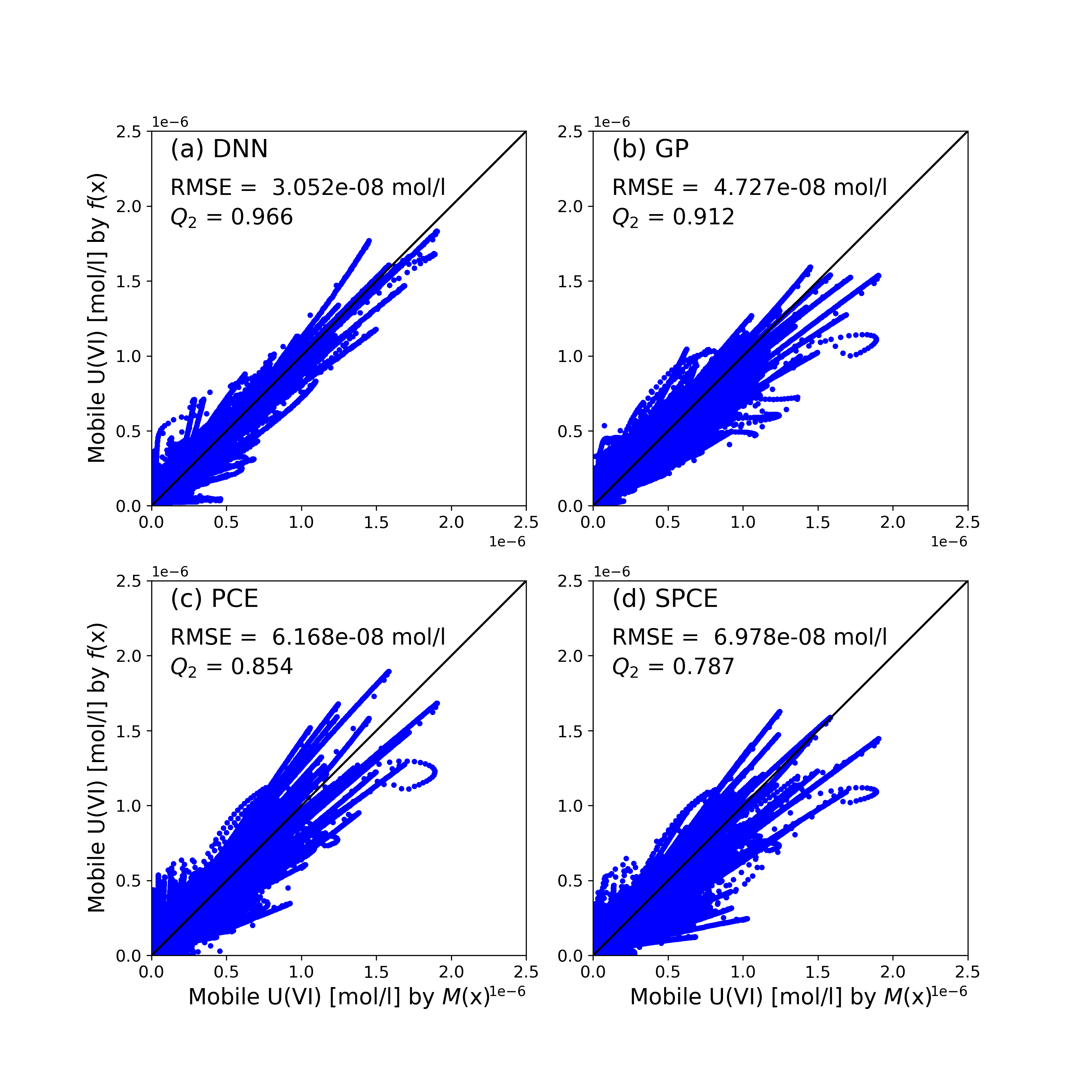}
	\caption{1-1 plots of emulation performance obtained for problem 2 when the emulators are built using 500 samples. The $x$-axis and $y$-axis present the true and emulated 240 $\times$ 1224 test data points, respectively. The RMSE and $Q_2$ coefficient denote the root-mean-square-error and coefficient of determination in testing mode, respectively, between the true and emulated 240 $\times$ 1224 test data points.}
	\label{fig8}
\end{figure}

\begin{figure}[H]
	\noindent\hspace{0cm}\includegraphics[width=40pc]{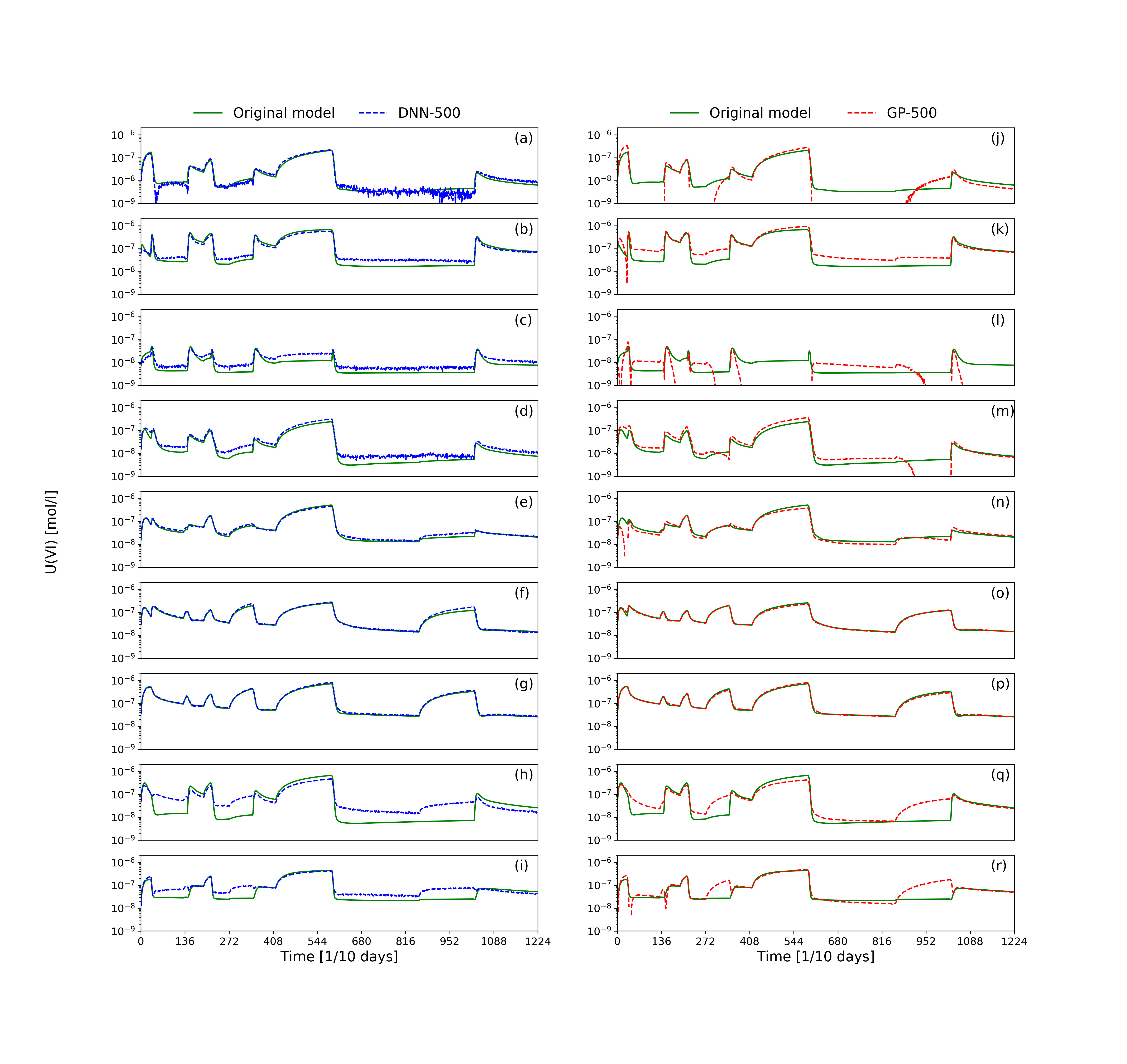}
	\caption{Original (green solid lines), DNN-emulated (left column, blue dashed lines) and GP-emulated (right column, red dashed lines) time series of 1224 U(VI) concentrations for problem 2. Each of the 9 rows corresponds to a different test example randomly chosen from the ensemble of 240 test examples. DNN-500 (blue dashed lines) signifies a DNN built using 500 training examples. GP-500 (red dashed lines) denotes a GP constructed using 500 training examples.}
	\label{fig9}
\end{figure}

The emulation results for the more complex problem 2 are not as good as for problem 1 (see Figures \ref{fig7}-\ref{fig9} and Table \ref{table6}). Also, the ranking of the emulation methods is globally similar to the ranking observed for problem 1. The only difference is that for the smallest training set size 75, the GP emulator now outperforms the DNN emulator (Figure \ref{fig7}). For training set sizes 175 and 500, the emulation accuracy reverses to DNN $>$ GP $>$ PCE $\approx$ sPCE (Figures \ref{fig8}-\ref{fig9} and Table \ref{table6}), just as for problem 1. Regarding the PCE and sPCE emulators they are again found to provide the worst accuracy while no clear difference in performance is noted between the two variants. Therefore, for this second problem as well the sPCE method is no longer considered in the next analyses.

\begin{table}[H]
	\caption{Emulation performance obtained for the 240-sample test set of problem 2 when the emulators are built using 175 samples. The RMSE and $Q_2$ coefficient denote the root-mean-square-error and coefficient of determination in testing mode, respectively, between the true and emulated 240 $\times$ 1224 test data points.}
	\begin{center}
		\begin{tabular}{lcc}%
			\hline
			Emulator & $Q_2$ & RMSE [mol/l] \\
			\hline
			DNN & 0.886 & 5.387 $\times$ 10$^{-8}$ \\
			GP & 0.848 & 5.954 $\times$ 10$^{-8}$  \\
			PCE & 0.654 & 8.711 $\times$ 10$^{-8}$  \\
			sPCE & 0.655 & 8.673 $\times$ 10$^{-8}$ \\
			\hline
		\end{tabular}
	\end{center}
	\label{table6}
\end{table}

\subsubsection{Sensitivity Analysis}
\label{res_pbm2_gsa}

\begin{figure}[H]
	\noindent\hspace{0cm}\includegraphics[width=35pc]{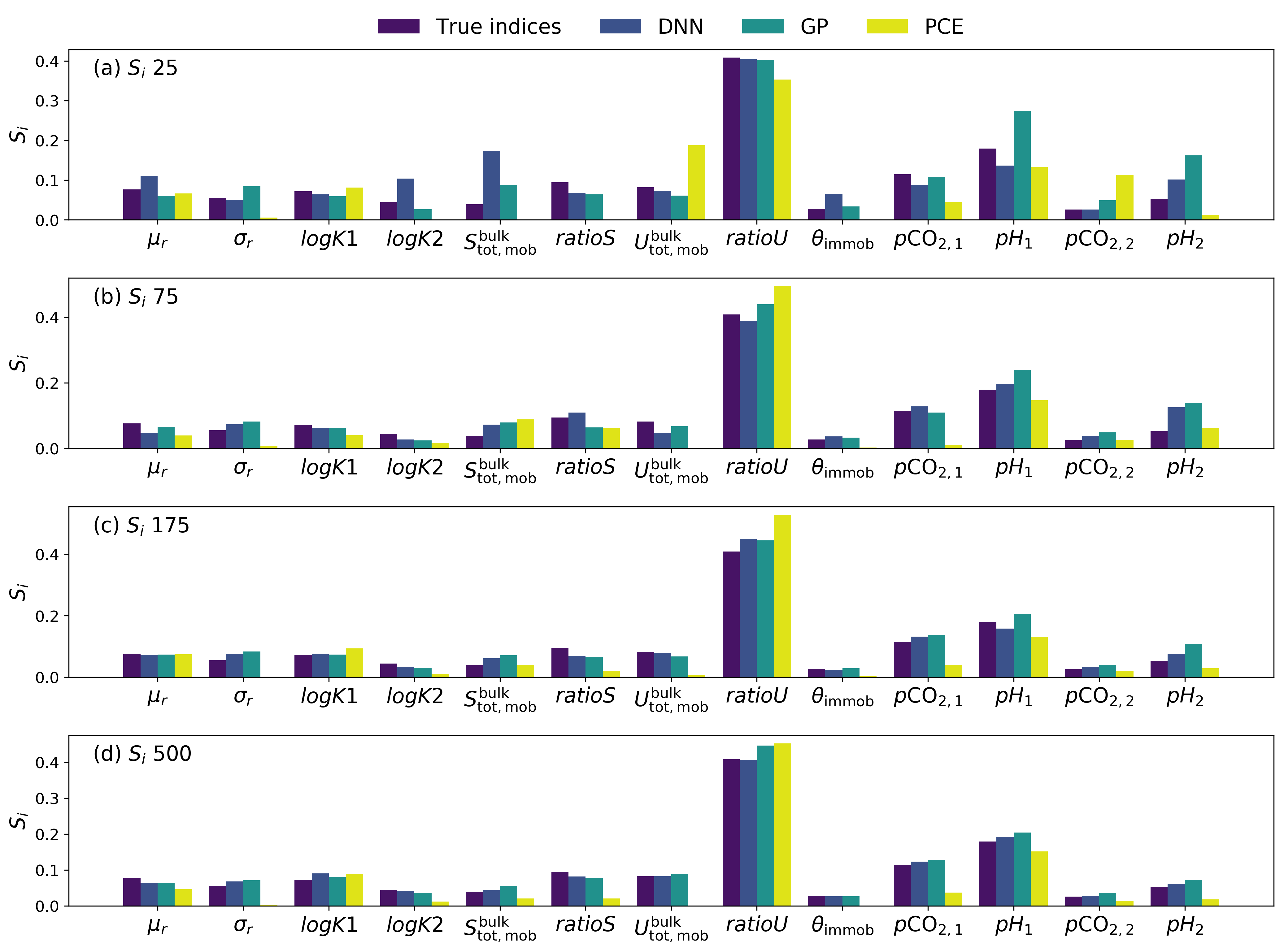}
	\caption{First ($S_i$) Sobol' sensitivity indices for the peak concentration associated with problem 2 as computed by the EASI method. $S_i$ 25, $S_i$ 75, $S_i$ 175 and $S_i$ 500 signify the emulated indices derived from using 25, 75, 175 and 500 training samples, respectively, to contruct the emulators. The true, DNN-based and GP-based indices are derived using the 240 points of the test set associated with problem 2. As commonly done with PCE, the PCE-based indices are obtained directly from manipulating the expansion terms. Yet feeding the PCE emulators with the 240 points to calculate the indices leads to very similar results (not shown).}
	\label{fig10}
\end{figure}

The first-order Sobol' sensitivity indices derived by the EASI approach, $S_i$, are displayed in Figure \ref{fig10} for the 13 model parameters of (1) the original RTM2 and (2) the 3 emulators constructed with 25, 75, 175 and 500 training exemples. For every method but PCE, the indices are calculated on the basis of the 240 parameter sets of the test set. As written earlier, the PCE-based indices are obtained directly from manipulating the expansion coefficients.

Not surprisingly, the estimated indices get better as the training set size increases. Furthermore, the DNN and GP emulators appear to perform equally well while the PCE-based indices are the least accurate ones. Indeed, for training set size 25 five parameters are erroneously assigned a zero $S_i$ by the PCE, and this is still the case for 3 model parameters: $\sigma_r$, $U^{\rm bulk}_{\rm tot,mob}$ and $\theta_{\rm immob}$, when the training set size is 500 (Figure \ref{fig10}).

\subsubsection{Uncertainty Propagation}
\label{res_pbm2_up}

Table \ref{table7} presents the true and emulated: (i) probability of exceeding the maximum concentration of 1.5 $\times$ 10$^{-6}$ mol/l, $p_{\rm MAX}$, (ii) mean peak concentration, $\mu_{\rm MAX}$ and (iii) standard deviation of the maximum concentration, $\sigma_{\rm MAX}$. The true and emulated values were all calculated over the 240 test samples used in section \ref{res_pbm2_gsa}. Here only the training set size 500 allows for a decent estimation by the DNN and GP emulators, which perform equally well for this training set size. In addition, for every training set size the PCE emulator again provides the worst approximation to $p_{\rm MAX}$.

\begin{table}[H]
	\caption{Uncertainty propagation results obtained for problem 2 from using the three emulators, DNN, GP and PCE, and three training set for emulator's construction: 75, 175 and 500.}
	\small{
		\begin{center}
			\begin{tabular}{cccccccccc}%
				\hline
				DNN-75 & GP-75 & PCE-75 & DNN-175 & GP-175 & PCE-175 & DNN-500 & GP-500 & PCE-500 \\
				\multicolumn{9}{c}{$\large{{p}_{\rm MAX}}$, true value is 0.0166}\\
				0.0207 & 0.0041 & 0.0156 & 0.0207 & 0 & 0 & 0.0166 & 0.0166 & 0.0250 \\
				\multicolumn{9}{c}{ }\\
				\multicolumn{9}{c}{$\large{\mu_{\rm MAX}}$ (10$^{-7}$ mol/l), true value is 5.346}\\
				5.066 & 5.029 & 5.176 & 5.048 & 5.000 & 5.064 & 5.357 & 5.125 & 5.211 \\
				\multicolumn{9}{c}{ }\\
				\multicolumn{9}{c}{$\large{\sigma_{\rm MAX}}$ (10$^{-7}$ mol/l), true value is 3.786}\\
				3.605 & 3.083 & 2.961 & 3.520 & 3.033 & 2.898 & 3.557 & 3.284& 3.531 \\
				\hline
			\end{tabular}
		\end{center}
	}
	\label{table7}
\end{table}

\section{Discussion}
\label{discussion}

In this paper, two increasingly complicated and CPU-demanding reactive transport modeling case studies are used to benchmark (state-of-the-art implementations of) 3 emulation methods: DNNs, GPs and PCE. For the considered computational effort allowed to training, we find that the DNN approach clearly outperforms PCE and is also superior to GP in reproducing the input - output behavior of the used 8-dimensional and 13-dimensional CPU-intensive RTMs. This even though the used computational budgets of 75, 175 and 500 forward runs are relatively small given the model parameter dimensionality of either 8 or 13. This is an important finding as DNNs are classically known to require a large training basis to be appropriately trained. Furthermore, the PCE and sparse PCE methods are shown to provide the worst emulation accuracy for every test case.

It follows from the above findings that for the considered case studies, uncertainty propagation is best achieved using a DNN emulator. The situation is slightly different for variance-based global sensitivity analysis. Here DNNs and GPs offer equally good approximations to the true first-order and total-order Sobol' sensitivity indices while again, PCE performs less well.

Most surprisingly, despite its outstanding performance for emulating the considered time series the DNN approach induces the largest bias in the inferred solution when used to replace the original RTM within a synthetic probabilistic model calibration. We believe that this unexpected behavior is at least partially due to the presence of a small deterministic noise in the DNN predictions. This noise appears to have a rather complex structure which can drive the emulated solutions far away from the true posterior distribution. When smoothed out to some extent using median or Savitzky-Golay filtering, the quality of the DNN-based posterior solutions increases but not up to a satisfying level. A potential solution to this drawback is to include a smoothing operator within the last layer of the DNN instead of performing smoothing as a post-processing step. In this way, the DNN weights and bias could be learned accounting for the final smoothing operation. We leave this idea for future work. 

For this inverse problem that involves 1224 measurement data corrupted with low noise and thus a peaky log-likelihood, the PCE approach is found to induce excessively large deviations between the true parameters and their emulated posterior distribution. Among the 3 emulation methods only the applied GP allows for finding emulated solutions that fit the high-quality measurement data to the appropriate noise level (log-likelihood value). Also, the GP-derived posterior distribution most closely match the true model parameters. The much better calibration performance of the GP emulator compared to the DNN emulator is likely due to its simpler interpolation engine that, for the fitted kernel and kernel parameters, induces smoother predictions. In addition, it is worth noting that sampling the original 8-dimensional RTM model with state-of-the-art MCMC for 10,000 iterations (which translates into a CPU-time of about 100 days) only permits exploration of a local mode of the posterior distribution. This highlights the complexity of the considered inverse problem.

Overall, our findings indicate that in the common case where the training set is relatively small (75 to 500 pairs of input - output examples) and fixed beforehand (no online construction), PCE is not the best emulation option for CPU-intensive RTMs. Rather one should use GPs or DNNs. Using a GP emulator seems to be the most robust choice as it performs well across all considered tasks: direct emulation, global sensitivity analysis, uncertainty propagation, and calibration. As stated above, our used fully-connected DNN can best emulate the original nonlinear RTMs but suffers from prediction errors that, although tiny, have a too complex structure for the DNN emulator to usefully replace the original model during calibration when the available measurement data are of high quality.

\section{Conclusion}
\label{conclusion}

We present a benchmark study of 3 methods for emulating CPU-intensive reactive transport models (RTMs): Gaussian processes (GPs), polynomial chaos expansion (PCE) and deep neural networks (DNNs). State-of-the-art open-source libraries are used for each emulation method while one forward RTM run incurs a CPU-time from about 1h (problem 1) to 1h30 - 5 days (problem 2). Besides direct emulation of the simulated time series, the increasingly computationally demanding tasks of global sensitivity analysis (GSA), uncertainty propagation and probabilistic calibration using Markov chain Monte Carlo (MCMC) are also scrutinized. The underlying idea is that the calculations can be made tractable by using a trained emulator in lieu of the original RTM. The DNN method is found to outperform both GPs and PCE in reproducing the input - output behavior of the used 8-dimensional and 13-dimensional CPU-intensive RTMs. This for relatively small training sample sizes used for constructing the emulators: from 75 to 500 samples. In addition, the two considered PCE variants, standard PCE and sparse PCE (sPCE), are found to always induce the least emulation accuracy. Accordingly, we find that for our test cases uncertainty propagation is best achieved using a DNN emulator. With respect to GSA, DNNs and GPs provide equally good approximations to the true first-order and total-order Sobol' sensitivity indices while PCE performs somewhat less well. Most surprisingly, despite its superior emulation capabilities the DNN method leads to the largest bias in the inferred solution when used to replace the original RTM within a synthetic probabilistic model calibration. This inverse problem involves 1224 measurement that are corrupted with low noise thereby causing the log-likelihood function to be rather peaky. The contradicting behavior of our used DNN is deemed to be at least partially caused by the small but complex deterministic noise that affects the DNN-based predictions. These prediction errors have such a complex structure that they can drive the emulated posterior distribution far away from the true one when the available measurement data are of high quality. Among the 3 emulators only the tested GP model allows for finding emulated posterior solutions that simultaneously (1) fit the high-quality measurement data to the appropriate noise level (log-likelihood value) and (2) most closely bracket the true model parameter values. Overall, our results highlight that when the available training set is relatively small (75 - to 500 input - output examples) and fixed beforehand, PCE is not the best choice for emulating CPU-intensive RTMs. Instead, GPs or DNNs are better options. Yet despite unrivaled emulation skills, our used DNN provides heavily biased calibration results. In contrast, using the GP method is a robust choice as it is found to perform relatively well across all considered test cases: direct emulation, global sensitivity analysis, uncertainty propagation, and calibration.

\section{acknowledgments}
All the codes used in this study are available from the first author (\url{elaloy@sckcen.be}).



\end{document}